\documentclass[aps,prd,preprintnumbers,showpacs,showkeys,nofootinbib,
superscriptaddress,fleqn,floatfix,tightenlines, 10pt]{revtex4-2}
\usepackage{amsmath,amsfonts,amssymb,amscd,amsxtra,amsthm}
\usepackage{graphicx}  % Include figure files
\usepackage{epstopdf}
\usepackage{dcolumn}  % Align table columns on decimal point 
\usepackage{bm}          % bold math
\usepackage{slashed}
\usepackage{cancel} 
\usepackage{mathtools}
\usepackage{amsbsy}
\usepackage{amstext}
\usepackage{threeparttable}

\usepackage[utf8]{inputenc}  
\usepackage{booktabs}
\usepackage[normalem]{ulem} % \sout{old text} for strikeout
\usepackage[dvipsnames,table,xcdraw]{xcolor} % For blue in-text
\usepackage{tabularx}
\usepackage{enumitem}  
\usepackage{array} 
\usepackage{slashed}
\usepackage{tikz}
\usepackage{float}
\usepackage{multirow}
\usepackage[symbol]{footmisc}
\usepackage{rotating}
\renewcommand\sout{\bgroup \color{red} \ULdepth=-.5ex \ULset}

\makeatletter
 
\begin{document}  
\preprint{INHA-NTG-01/2022}
\title{Axial-vector transition form factors of the baryon octet to 
the baryon decuplet with flavor SU(3) symmetry breaking}
%--------------------------------------------------
%--------------------------------------------------
\author{Jung-Min Suh}
\email[E-mail: ]{suhjungmin@inha.edu}
\affiliation{Department of Physics, Inha University, Incheon 22212,
Republic of Korea}

\author{Yu-Son Jun}
\email[E-mail: ]{ysjun@inha.edu}
\affiliation{Department of Physics, Inha University, Incheon 22212,
Republic of Korea}

\author{Hyun-Chul Kim}
\email[E-mail: ]{hchkim@inha.ac.kr}
\affiliation{Department of Physics, Inha University, Incheon 22212,
Republic of Korea}
\affiliation{School of Physics, Korea Institute for Advanced Study
(KIAS), Seoul 02455, Republic of Korea}
%--------------------------------------------------
\date{\today}
\begin{abstract} 
We investigate the axial-vector transition form factors of the baryon
octet to the baryon decuplet within the framework of the chiral
quark-soliton model, with the effects of flavor SU(3) symmetry
breaking included. We consider the rotational $1/N_c$ corrections and  
regard the strange current quark mass as a perturbation. We compare
the present results for the $\Delta \to N$ axial-vector transition
with those from other models and lattice QCD. We also compute all
possible axial-vector transitions from the baryon decuplet to the
octet with the strangeness changed, i.e., $|\Delta S|=1$. We obtain 
the value of the essential form factor $C_5^A$ for the $\Delta
\to N$ axial-vector transition at the zero momentum transfer
($Q^2=0$). Furthermore, the present results are in good agreement with
those fitted with the T2K data. We extract the value of the
axial-vector mass $M_A$ compared to the data.    
\end{abstract}
\maketitle
%--------------------------------------------------
\section{Introduction}
The axial-vector transitions of SU(3) baryons address multi-faceted
issues on strong and weak processes of hadrons. A typical axial-vector
transition can be found in hyperon semileptonic 
decays (HSD)~\cite{Garcia:1985xz,Cabibbo:2003cu}. While most of the
axial-vector transition constants for the baryon octet HSD were known 
experimentally~\cite{PDG}, experimental evidence for the $\Xi^-
\to \Xi^0e^{-}\bar{\nu}_e$ decay is still
elusive~\cite{BESIII:2021emv}. HSDs provide information on
the Cabibbo-Kobayashi-Maskawa (CKM) mixing matrix elements $|V_{ud}|$
and $|V_{us}|$~\cite{C,KM} in addition to the pion and kaon 
decays~\cite{Gamiz:2004ar, Pich:2013lsa, Seng:2018yzq,
  Czarnecki:2019iwz}. While the CKM mixing angles extracted from HSDs
can only play an auxiliary role, it is still of great importance to
determine the unitarity of the CKM matrix:
$|V_{ud}|^2+|V_{us}|^2+|V_{ub}|^2=1$~\cite{Cabibbo:2003ea, 
  Sharma:2010zza, Garcia:1991pu}. HSDs also cast light on the
structure of the SU(3) baryons. The experimental data on the
semileptonic decay constants reveal a certain pattern of explicit
flavor SU(3) symmetry breaking~\cite{Donoghue:1986th, Roos:1990en, 
Ratcliffe:1998su, Flores-Mendieta:1998tfv, Yang:2015era,
Wang:2019alu}. The baryon decuplet, on the other hand, decays in the
baryon octet primarily through the strong interaction except for the
$\Omega^-$ baryon. Nevertheless, understanding the $\Delta\to N$
axial-vector transition form factor is critical 
because it provides crucial information for describing the weak single
pion production ($\nu_\mu p \to \mu^- \pi^+  
p$) from neutrino-nucleon scattering~\cite{Adler:1968tw,
Barish:1978pj, Radecky:1981fn, Kitagaki:1986ct, Kitagaki:1990vs}. Since 
neutrino-nucleon scattering holds an essential clue on the neutrino
oscillations, there has been a great deal of experimental 
programs such as the No$\nu$a, MiniBooNE, T2K, NusTEC, Miner$\nu$a,
DUNE, and SND@LHC experiments~\cite{Nova, MiniBooNE:2010bsu,
  T2K:2014xyt, T2K:2021xwb, NuSTEC:2017hzk, MINERvA:2013bcy,
  Lovato:2020kba, SND} in higher energy regions (see also a recent
review~\cite{Mosel:2016cwa}). Very energetic neutrinos in future
experiments such as the DUNE and SND@LHC will be available, so one can
have a possible opportunity to study the structure of strange baryons
in neutrino-nucleon scattering. These experiments will shed light on
the axial-vector structure of nonstrange and strange 
baryon resonances. In addition, Alexandrou et al. reported the results 
on the $\Delta\to N$ axial-vector transition form factors based on
lattice QCD~\cite{Alexandrou:2007eyf, Alexandrou:2010uk}. Thus, it is
of great interest to scrutinize the axial-vector transitions from the
baryon decuplet to the octet, which will give multiple perspectives on
the structure of baryons. 

There have already been many theoretical works on the axial-vector
transition form factors for the nucleon to the $\Delta$ excitation: 
for example, the relativistic quark
model (RQM)~\cite{Ravndal:1973xx,LeYaouanc:1976ea,Korner:1977rb}
, the isobar model (IM)~\cite{Fogli:1979cz}, the nonrelativistic
quark model (NRQM)~\cite{Hemmert:1994ky, Liu:1995bu}, the linear
$\sigma$ model (LSM) and the cloudy bag model
(CBM)~\cite{Golli:2002wy}, the chiral 
constituent quark model ($\chi$CQM)~\cite{Barquilla-Cano:2007vds},
baryon chiral perturbation theory~\cite{Zhu:2002kh, Geng:2008, Yao:2018pzc,
  Unal:2021byi}, the Barbero-Lopez-Mariano model
(BLM)~\cite{Barbero:2008zza,Barbero:2013eqa}, the $\Delta$-pole
dominance model~\cite{Hernandez:2010bx}, the light-cone QCD sum rule
(LCSR)~\cite{Kucukarslan:2015urd} and the nonlinear $\sigma$ model
(NLSM)~\cite{Alvarez-Ruso:2015eva}.  
The $\Delta\to N$ axial-vector transition form factor has often been
parametrized either by the dipole-type form factor or by Adler's
parametrization~\cite{Adler:1968tw}. These parametrizations being  
used, the value of the axial transition mass $M_A$ for the $N\to \Delta$
axial-vector transition can be extracted from the experimental
data. Based on the ANL~\cite{Barish:1978pj, Radecky:1981fn} and BNL 
data~\cite{Kitagaki:1986ct, Kitagaki:1990vs}, many theoretical and
experimental efforts were put on extracting the values of the
$\Delta\to N$ axial vector form factor $C_5^A(0)$ and $M_A$: the ranges
of their values lie in $0.8-1.2$ and $0.8-1.0$ GeV,
respectively~\cite{Barish:1978pj, Radecky:1981fn,
  Barbero:2008zza,Barbero:2013eqa, Rein:1980wg,
  Lalakulich:2005cs, Aachen-Bonn-CERN-Munich-Oxford:1980iaz}.    
The off-diagonal Goldberger-Treiman (GT) relation for the 
$\Delta\to N$ axial-vector transition constant predicts $C_5^A(0)$ to
be around $1.2$~\cite{Graczyk:2009qm,Graczyk:2021oyl} with the
experimental data on the $\Delta$ decay width considered.
In this context, the deviation of the off-diagonal GT relation was
also discussed and was found small~\cite{Zhu:2002kh, Bernard:2001rs}.
However, Ref.~\cite{Hernandez:2007qq} found the smaller value
$C_5^A=0.87\pm 0.08$, which is more reliable for MiniBooNE and T2K
experiments. In addition, the nucleon-nucleon potential such as the
Bonn-J\"ulich potential~\cite{Holzenkamp:1989tq} takes the smaller
value of the $\pi N\Delta$ coupling constant ($f_{\pi
  N\Delta}^2/4\pi=0.224$) than that derived from the $\Delta$ decay
width ($f_{\pi N\Delta}^2/4\pi=0.36$). If one uses this smaller value
of $f_{\pi  N\Delta}$, one would get a smaller value of $C_5^A$ from
the off-diagonal GT relation. 

We also want to mention that there are only a few studies on the
axial-vector transitions from the baryon decuplet to the octet. The
axial-vector transition constants from the baryon decuplet to the
octet with the strangeness conserved ($\Delta S=0$) were already
computed within the chiral quark-soliton model
($\chi$QSM)~\cite{Ledwig:2008rw}. Those with $|\Delta S|=1$ were
investigated in a pion mean-field approach, where all possible
parameters were fixed by using the experimental data on
HSD~\cite{Yang:2018idi}. In the present work, we will extend the 
previous study to compute all possible axial-vector transition form
factors from the baryon decuplet to the octet up to a momentum
transfer $Q^2 \le 1\,\mathrm{GeV}^2$ within the framework of the
self-consistent $\chi$QSM with explicit flavor SU(3) symmetry breaking
considered.    

The $\chi$QSM is a pion mean-field approach~\cite{Diakonov:1987ty,
  Wakamatsu:1990ud, Diakonov:1997sj}. As Witten
proposed~\cite{Witten:1979kh, Witten:1983tx}, a baryon in the large
$N_c$ (the number of colors) limit emerges as a state consisting of
$N_c$ valence quarks, bound by the pion mean field, since the mesonic
quantum fluctuations are suppressed by $1/N_c$. The pion mean field
arises from the classical solution of the equation of motion, which
can be solved self-consistently. This procedure is nothing but a 
Hartree approximation~\cite{Kahana:1984be} (see also a
review~\cite{Christov:1995vm}). The presence of the $N_c$ valence
quarks makes the Dirac continuum polarized, which creates the pion
mean field by which the $N_c$ valence quarks are bound. Recently,
it was shown that this mean-field approach could also describe singly
heavy baryons, i.e., a singly heavy baryon as a bound state of 
the $N_c-1$ valence quarks~\cite{Yang:2016qdz} (see also a recent
review~\cite{Kim:2018cxv}). The classical solution obtained by this
self-consistent procedure is called the classical nucleon or the
chiral soliton, which needs to be quantized. While we ignore the
$1/N_c$ mesonic quantum fluctuations, we have to deal with the zero
modes related to continuous translational and rotational
symmetries. Since we will compute form factors of the SU(3) baryons,
we have to consider the translational zero modes, which will yield
the Fourier transforms, and the rotational zero modes in SU(3), with
the SU(2) soliton embedded into SU(3). This embedding preserves
the hedgehog symmetry of the SU(2) soliton. 
Assuming that the angular velocity of the soliton is slow and the
mass of the strange current quark ($m_s$) is small, we will treat them
as perturbations. Thus, we will consider the rotational $1/N_c$
and linear $m_s$ corrections. 
The model has been successfully applied to various observables of the
SU(3) baryons: 
for example, the electromagnetic
structures~\cite{Kim:1995mr,Kim:1995ha,Wakamatsu:1996xm, Kim:1997ip, 
  Silva:2013laa}, strange form factors ~\cite{Kim:1995hu,
  Silva:2001st, Silva:2005qm, Kim:1995bq, Kim:1996vk, Kim:2019gka,
  Kim:2020lgp}, axial-vector form factors~\cite{Silva:2005fa,
  Jun:2020lfx}, tensor charges and corresponding form
factors~\cite{Kim:1995bq, Kim:1996vk, Ledwig:2010tu, Ledwig:2010zq}, 
semileptonic decays~\cite{Kim:1997ts,Ledwig:2008ku,Yang:2015era},
radiative transition~\cite{Watabe:1995xy,Silva:1999nz,Ledwig:2008zrm},
the nucleon parton distributions~\cite{Diakonov:1996sr,Diakonov:1997vc,
Pobylitsa:1996rs, Goeke:2000wv,Schweitzer:2001sr,Pobylitsa:1998tk},   
and the gravitational form factors~\cite{Goeke:2007fp, Kim:2020nug} of
the nucleon. In this work, we will concentrate on all possible
axial-vector transitions from the baryon decuplet to the baryon octet,
including both $\Delta S=0$ and $|\Delta S| =1$ transitions.  

The present work is organized as follows: In Section II, we define the
axial-vector transition form factors from the baryon decuplet to the
baryon octet, which parametrize the matrix elements of the
axial-vector current. In Section III, we briefly review the formalism
of the $\chi$QSM in the context of the derivation of the axial-vector 
transition form factors. In Section IV, we present their numerical
results. We first discuss the effects of
flavor SU(3) symmetry breaking. We then compare the numerical
results with those from the lattice data. We present the results for
the axial-vector transition constants and compare them with those from
other theoretical works. We also provide the results for the
transition radii and dipole mass that will be useful for describing
hadronic processes. In the last section, we summarize 
the present work and draw conclusions.  
%=================
\section{Axial-vector transition form factors from
  the baryon octet to the baryon decuplet} 
\label{sec:2}
The axial-vector current is defined as
\begin{align}
A^{\chi}_\mu (x) = \bar{\psi} (x) \gamma_\mu \gamma_{5} 
\frac{\lambda^{\chi}}{2} \psi(x),
\label{eq:AxialCUR}
\end{align}
where $\lambda^{\chi}$ are the short-handed notation for the flavor
SU(3) Gell-Mann matrices: for the strangeness-conserving ($\Delta
S=0$ or $\chi=3,\,1\pm i2$) transitions and strangeness-changing ones
($|\Delta S|=1$ or $\chi=4 \pm i5$), we define $\lambda^{1\pm i2}$ and
$\lambda^{4\pm i5}$ respectively by 
\begin{align}
\lambda ^{1\pm i2} = \frac1{\sqrt{2}}(\lambda^{1}\pm i\lambda^{2}),\;\;\;
\lambda ^{4\pm i5} = \frac1{\sqrt{2}}(\lambda^{4}\pm i\lambda^{5}).
\end{align}
$\psi(x)$ stands for the quark field $\psi=(u,d,s)$. Since we deal
with the baryon decuplet, 
the Lorentz structure of the spin-3/2 baryons should be
considered~\cite{Adler:1968tw}. This means that 
we have more form factors than the case of the baryon octet, which are
often called the Adler form factors. Then the matrix element of the 
axial-vector current between the baryon decuplet and the baryon octet
can be parametrized in terms of four different real form
factors~\cite{LlewellynSmith:1971uhs}: 
\begin{align}
\langle B_{8}(p_{8},J_3') | A_\mu^{\chi}(0) | B_{10}(p_{10},J_3) \rangle 
&=  \overline{u}(p_{8},J_3') \left[  \left 
  \{\frac{C_{3}^{A(\chi)}(q^2)}{M_{8}} \gamma^{\nu} +
  \frac{C_{4}^{A(\chi)}(q^2)}{M_{8}^{2}}p_{10}^{\nu}\right \} 
  (g_{\alpha \mu}g_{\rho \nu}-g_{\alpha \rho}g_{\mu \nu})q^{\rho}
  \right. \cr
&\hspace{2cm} \left. + \,C_{5}^{A(\chi)}(q^2)g_{\alpha \mu}
  +\frac{C_{6}^{A(\chi)}(q^2)}{M_{8}^2}q_{\alpha}q_{\mu} 
   \right ]  u^{\alpha}(p_{10},J_3), 
\label{eq:MatrixEl1}
\end{align}
where $M_{8}$ and $M_{10}$ designate respectively the masses of the
baryon octet and decuplet. $g_{\alpha \beta}$ denote the metric 
tensor of Minkowski space, expressed as 
$g_{\alpha \beta} =\mathrm{diag}(1,\,-1,\,-1,\,-1)$.
In the rest frame of a decuplet baryon, $p_{10}^{\alpha}$,
$p_{8}^{\alpha}$ and $q^{\alpha}$ represent respectively the momenta
of a decuplet baryon, that of an octet baryon and the momentum
transfer, which are written by
\begin{align}
&p_{10}=(M_{10},\bm{0}),\;\;\; p_{8}=(E_{8},-\bm{q}), \;\;\;
  q=(\omega_{q},\bm{q}) 
\end{align}
with $q^2=-Q^2>0$. 
Thus, the three-vector momentum and energy of the momentum transfer
are given as  
\begin{align}
&|\vec{q}|^{2} = \left(\frac{M_{10}^{2}+M_{8}^{2}+Q^{2}}{2M_{10}}
  \right)^{2}-M_{8}^{2}    \cr  
&\omega_{q}=\left(\frac{M_{10}^{2}-M_{8}^{2}-Q^{2}}{2M_{10}} \right). 
\end{align}
$u^\alpha (p_{10},\,J_3)$ stands for the Rarita-Schwinger
spinor that describes a decuplet baryon with spin
3/2, carrying the momentum $p_{10}$ and spin $J_3$. It 
can be expressed by the combination of the polarization vector and the 
Dirac spinor, $u^\alpha (p_{10},\,J_3)= 
\sum_{i,s} C^{\frac{3}{2} J_3}_{1i\,\frac{1}{2}s} \epsilon^{\alpha}_{i}
(p_{10}) u_{s}(p_{10})$. It satisfies the Dirac equation and the auxiliary 
equations $p_{10 \alpha} u^\alpha(p_{10},J_3)=0$ and $\gamma_\alpha 
u^\alpha(p_{10},J_3)=0$~\cite{Rarita:1941mf}. $u(p_{8},J_3')$ denotes
the Dirac spinor for an octet baryon.  

In the current work, we will concentrate on $C_5^A(q^2)$. 
The transition matrix element of the axial-vector current is involved
in the cross section of neutrino-nucleon scattering. As discussed in
many references (for example, see
Refs.~\cite{LlewellynSmith:1971uhs,Barbero:2013eqa}), all other 
terms except for $C_5^A$ are suppressed by the ratio $q/M_N$ or
$q^2/M_N^2$ in the case of neutrino quasi-elastic scattering. Thus,
the value of $C_5^A(0)$ can be extracted from the neutrino
scattering data. Moreover, $C_5^A(0)$ is directly connected to the
strong $\pi N\Delta$ coupling constant with the Goldberger-Treiman
relation~\cite{Goldberger:1958tr,Goldberger:1958vp, Nambu:1960xd}.  
 The divergence of the axial-vector current should vanish in 
the chiral limit~\cite{Alfaro}
\begin{align}
i \bar{u}_{B_8}(p_8',J_3') q_\mu[C_5^A(q^2) + C_6^A(q^2)
  \frac{q^2}{M_{8}^{2}}]   u_{B_{10}}^\mu(p_{10},J_3), 
\end{align}
which yields
\begin{align}
C_5^A(q^2) + C_6^A(q^2) \frac{q^2}{M_{8}^{2}} =0.  
\end{align}
This indicates that $C_6^A(q^2)$ must have a pole at $q^2=0$ because
$C_5^A(0)$ does not vanish. The pole term $C_6^A(q^2)$ leads to the
following structure 
\begin{align}
\bar{u}_{B_8} (p_8',J_3') q_\mu C_6^A(q^2) u_{B_{10}}^\mu(p_{10},J_3) 
\rightarrow f_\pi \frac{g_{B_8B_{10} M}}{M_8+M_{10}} \bar{u}_{B_8} (p_8',J_3')
  q^\mu g_{\mu\nu} \frac{i}{q^2} 
  u_{B_{10}}^\nu(p_{10},J_3) ,   
\end{align}
where $g_{B_8B_{10} M}$ denotes the strong coupling constant for a
vertex with decuplet and octet baryons, and an octet meson.
Using this relation, we find
\begin{align}
\lim_{q^2\to 0} \left(C_5^A (q^2) +
  C_6^A(q^2)\frac{q^2}{M_{8}^{2}}\right) = 
  \lim_{q^2\to 0} \left[C_5^A(q^2) - \frac{g_{B_8B_{10}
  M}}{M_8+M_{10}} f_\pi \right] = 0, 
\end{align}
which gives the well-known Goldberger-Treiman relation (GTR) for a
spin-3/2 baryon 
\begin{align}
  C_5^A(0) = f_\pi \frac{g_{B_8B_{10} M}}{M_8+M_{10}}.
\end{align}
The meson-baryon strong coupling constants have been already
investigated in this pion mean-field approach, where all dynamical
parameters were fixed by the experimental data on
HSDs~\cite{Yang:2018idi}. We want to mention that the GTR has a
certain discrepancy~\cite{Pagels:1969ne}.

The form factors  $C_{5}^{A, 10 \rightarrow 8}(q^{2})$ are determined
by the transition matrix elements of the spatial component of the
axial-vector current
\begin{align}
C_{5}^{A, 10 \rightarrow 8}(q^{2})&=
   \frac{\sqrt{3M_{8}}}{\sqrt{E_{8}+M_{8}}} 
\Bigg[\int \frac{d\Omega_{q}}{4\pi} \langle
 B_{8}(p_{8},J_3)|\bm{e}_{0}\cdot 
\bm{A}|B_{10}(p_{10},J_3)\rangle \cr  
&-\sqrt{5\pi} \int \frac{d\Omega_{q}}{4\pi}
Y_{20}(\Omega_{q})\langle B_{8}(p_{8},J_3)|
\bm{e}_{0}\cdot \bm{A}|B_{10}(p_{10},J_3)\rangle  \Bigg], 
\label{eq:extC5}
\end{align}
where $\bm{e}_{0}$ denotes the polarization vector in the spherical
basis, i.e. $\bm{e}_{0}=(0,0,1)$ and $\bm{A}$ stands for the spatial
component in the vector form: $\bm{A} = \bar{\psi}(x)\bm{\gamma}
\gamma_{5}\frac{\lambda^{\chi}}{2}\psi(x)$.  
We fix the third component of the spin states for the
baryon octet and decuplet to be $J_3=1/2$ for convenience. We will now 
compute these transition matrix elements in the present work.  

%%%%%%%%%%%%%%%%%%%%%%%%%%
\section{Axial-vector transition form factors in the chiral quark-soliton model} 
%%%%%%%%%%%%%%%%%%%%%%%%%%
\label{sec:3}
The SU(3) $\chi$QSM starts from the low-energy effective partition
function in Euclidean space  
\begin{align}
\mathcal{Z}_{\chi \mathrm{QSM}}&=\int \mathcal{D}\psi^a \mathcal{D}
                                 \psi^{\dagger} \mathcal{D}\pi^a 
 \exp{\Big[-\int d^{4}x \psi^{\dagger}  iD(\pi)\psi\Big]}  
=\int D\pi \exp{(-S_{\mathrm{eff}})},
\label{eq:partitionf} 
\end{align}
where $\psi$ and $\pi^a$ represent the quark and
pseudo-Nambu-Goldstone boson fields (pNG). The $S_{\mathrm{eff}}$ is the
effective chiral action expressed as 
\begin{align}
S_{\mathrm{eff}}[\pi^{a}] \;=\; -N_{c}\mathrm{Tr}\ln D, 
\label{eq:echa}
\end{align}
where $\mathrm{Tr}$ stands for the functional trace running over
spacetime and all relevant internal spaces. 
The $N_{c}$ is the number of colors, and $D(U)$ designates the Dirac
differential operator defined by 
\begin{align}
D := i\slashed{\partial} + i M U^{\gamma_5} + i \hat{m},
\label{eq:Dirac}
\end{align}
where $M$ denotes the dynamical quark mass. Note that $M$ is
originally momentum dependent, which comes from the instanton vacuum.  
The momentum-dependent dynamical quark mass is originated from the
quark zero mode in the presence of the
instanton~\cite{Diakonov:1985eg, Diakonov:2002fq}. Since we use the
constant dynamical quark mass in the present work, we have to
introduce the regularization to tame the divergence of the quark
loops. $U^{\gamma_5}(x)$ in Eq.~\eqref{eq:Dirac} represents the 
SU(3) chiral field defined by  
\begin{align}
U^{\gamma_5}(x) := \frac{1+\gamma_5}{2} U(x) + \frac{1-\gamma_5}{2}
  U^\dagger (x)
\end{align}
with $U(x) = \exp(i\lambda^a \pi^a (x) /f_\pi)$. $f_\pi$ is the scale
factor that will be identified as the pion decay constant. 
$\hat{m}$ in Eq.~\eqref{eq:Dirac} represents the current quark mass
matrix given as  
$\hat{m}=\mathrm{diag}(m_{\mathrm{u}},\, m_{\mathrm{d}},\,
m_{\mathrm{s}})$ in flavor space. 
We assume isospin symmetry in this work, so that the current quark
masses of the up and down quarks are set equal to each other,
i.e. $m_{\mathrm{u}}=m_{\mathrm{d}}$ with their average mass 
$\overline{m}=(m_{\mathrm{u}} + m_{\mathrm{d}})/2$. Then, the current 
quark mass matrix is written as $\hat{m} =
\mathrm{diag}(\overline{m},\, \overline{m},\, m_{\mathrm{s}}) =
\overline{m} +\delta m$.
$\delta m$ includes the mass of the strange current quark, which can be 
decomposed as 
\begin{align}
\delta m \;=\; m_{1} \bm{1} + m_{8} \lambda^{8}. 
\label{eq:deltam}
\end{align}
$m_1$ and $m_8$ denote the singlet and octet components of 
the current quark masses respectively:
$m_1=(-\overline{m}+m_{\mathrm{s}})/3$ and
$m_8=(\overline{m}-m_{\mathrm{s}})/\sqrt{3}$. 
The Dirac operator~\eqref{eq:Dirac} with $\gamma_4$ can be written as
\begin{align}
\gamma_4 D = -i\partial_{4} + h(U(\pi^{a})) - \delta m,
\end{align}
where $\partial_4$ stands for the time derivative in Euclidean space. 
$h(U)$ is called the one-body Dirac Hamiltonian written as 
\begin{align}
h(U) \;=\;
i\gamma_{4}\gamma_{i}\partial_{i}-\gamma_{4}MU^{\gamma_{5}} -
\gamma_{4} \overline{m}\, .
\label{eq:DiracHam}  
\end{align}

As mentioned previously, the pion mean field arises as the
solution of the classical equation of motion, which is derived from 
$\delta S_{\mathrm{eff}}/\delta P(r) =0$. The equation of motion can
be solved self-consistently, which resembles the Hartree approximation
in many-body problems. In solving the classical equation of motion or
minimzing the classical nucleon mass, one needs to find the pion field
with proper symmetry. In flavor SU(2), three components of the
pion field are coupled to three dimensional space, so that the pion
fields are expressed in terms of the profile function $P(r)$ for the
chiral soliton  
\begin{align}
\pi^i = n^i P(r), \; i=1,\,2,\,3,
\end{align}
where $n^i=x^i/r$ with $r=|\bm{x}|$. This expression is often called
the hedgehog ansatz and the corresponding symmetry is known to be
hedgehog symmetry. Since we want to keep this hedgehog symmetry of the
pion field preserved~\cite{Pauli:1942kwa, Witten:1983tx} also in
SU(3), we embed the SU(2) $U_{\mathrm{SU(2)}}(x)$ field into SU(3).
The SU(3) $U(x)$ field can be constructed by the trivial 
embedding~\cite{Witten:1983tx} 
\begin{align}
U(x) = \exp(i\pi^a \lambda^a/f_\pi) = 
  \begin{pmatrix}
    \exp(i\bm{n}\cdot \bm{\tau}P(r)/f_\pi) & 0 \\ 0 & 1
  \end{pmatrix},
\end{align}
where $\pi^a$ are set equal to zero for $a=4,\cdots 8$. The zero-mode
quantization with this embedding will correctly yield the spectrum of
the SU(3) baryons. 

We can compute the matrix elements of the axial-vector
current~\eqref{eq:MatrixEl1} by using the functional integral 
\begin{align}
&\langle B(p',\, J_3') | A^{a}_\mu(0) |B(p,\,J_3)\rangle =
  \frac1{\mathcal{Z}_{\chi \mathrm{QSM}}} \lim_{T\to\infty}
  \exp\left(i p_4\frac{T}{2}  
  - i p_4' \frac{T}{2}\right) \int d^3x d^3y \exp(-i \bm{p}'\cdot 
  \bm{y} + i \bm{p}\cdot \bm{x}) \cr
&\hspace{1cm}\times \int \mathcal{D} \pi^a \int \mathcal{D} 
  \psi \int \mathcal{D} \psi^\dagger 
J_{B}(\bm{y},\,T/2) \psi^\dagger(0) 
  \gamma_4\gamma_\mu \gamma_{5} \frac{\lambda^{a}}{2} \psi(0) 
  J_B^\dagger (\bm{x},\,-T/2) \exp\left[-\int d^4 r \psi^\dagger 
  iD(\pi^a) \psi\right],
\label{eq:correlftn}
\end{align}
where the baryon states $|B(p,\,J_3)\rangle$ and $\langle
B(p',\,J_3')|$ are respectively written as
\begin{align}
|B (p,\,J_3)\rangle &= \lim_{x_4\to-\infty} \exp(i p_4 x_4)
  \frac1{\sqrt{\mathcal{Z}_{\chi \mathrm{QSM}}}} \int d^3 x
                      \exp(i\bm{p}\cdot \bm{x})  
  J_B^\dagger (\bm{x},\,x_4)|0\rangle,\cr
\langle B(p',\,J_3')| &= \lim_{y_4\to\infty} \exp(-i p_4' y_4)
  \frac1{\sqrt{\mathcal{Z}_{\chi \mathrm{QSM}}}} \int d^3 y
                        \exp(-i\bm{p}'\cdot \bm{y})  
  \langle 0| J_B (\bm{y},\,y_4).
\label{eq:correl}
\end{align}
Here, $J_B(x)$ represents the Ioffe-type current that consists of the
$N_c$ valence quarks~\cite{Ioffe:1981kw, Diakonov:1987ty}  
\begin{align}
J_B(x) = \frac1{N_c!} \epsilon_{i_1\cdots i_{N_c}} \Gamma_{JJ_3
  TT_3 Y}^{\alpha_1\cdots \alpha_{N_c}} \psi_{\alpha_1 i_1} (x)
  \cdots \psi_{\alpha_{N_c} i_{N_c}}(x),  
\end{align}
with spin-flavor and color indices $\alpha_1\cdots \alpha_{N_c}$
and $i_1\cdots i_{N_c}$, respectively. The matrices
$\Gamma_{JJ_3 TT_3 Y}^{\alpha_1\cdots \alpha_{N_c}}$ carry the spin
and flavor quantum numbers of the baryon, i.e.,
$JJ_3TT_3Y$. Similarly, we can express the creation current operator 
$J_B^\dagger(x)$~\cite{Diakonov:1987ty, Christov:1995vm}.

To quantize the chiral soliton, we have to perform the
functional integral over the pNG fields. Since we use the pion
mean-field approximation or the saddle-point approximation, we neglect
the $1/N_c$ quantum fluctuations of the pNG fields or the pion-loop
corrections. However, we have to take into 
account the zero modes completely, which do not change the energy of
the soliton. Thus, the functional integral over the $U$ field is replaced
by rotational and translational zero modes that are written as 
\begin{align}
\tilde{U}(\bm{r}, t) = A(t) U(\bm{r} -\bm{Z}(t)) A^\dagger (t),
\label{eq:25}
\end{align}
where $A(t)$ belongs to an SU(3) unitary matrix and $\bm{Z}(t)$
correspond to the translational zero modes. The Dirac operator in
Eq.~\eqref{eq:Dirac} is then changed as 
\begin{align}
\tilde{D} = \partial_4 + h(U) + A^\dagger(t) \dot{A}(t) - i\gamma_4
  \dot{\bm{Z}} \cdot \bm{\nabla} + \gamma_4 A^\dagger (t) (\delta m)
  A(t),    
\end{align}
where $A^\dagger(t) \dot{A}(t)$ is the angular velocity of the soliton
 $\Omega(t)$ in Euclidean space
\begin{align}
A^\dagger(t) \dot{A}(t) = i\Omega = \frac12 i\Omega^a \lambda^a    
\end{align}
and $\dot{\bm{Z}}$ designates the translational velocity of the
soliton 
\begin{align}
\dot{\bm{Z}} = \frac{d\bm{Z}}{dt}. 
\end{align}
Then the effective action under the zero-mode quantization is
expressed as 
\begin{align}
\tilde{S}_{\mathrm{eff}} = -N_c \mathrm{Tr}\ln \left[ \partial_4 +
  A^\dagger (t) \dot{A}(t) - i\gamma_4 \dot{\bm{Z}} \cdot \bm{\nabla} +
  \gamma_4 A^\dagger (t) (\delta m) A(t) - \gamma_4 a_\mu
  \gamma_\mu\gamma_5  A^\dagger   (t) \lambda^\chi A(t)\right],   
\end{align}
where $a_\mu$ stands for the external axial-vector source field. 
Expanding the zero-mode quantized effective action in powers of
angular and translational velocities that are proportional to $1/N_c$,
we obtain the action as
\begin{align}
\tilde{S}_{\mathrm{eff}} \approx -N_c \mathrm{Tr}\ln D +
  S_{\mathrm{rot}}[A] + S_{\mathrm{trans}}[\bm{Z}], 
\end{align}
where
\begin{align}
S_{\mathrm{rot}}[A] = \frac12 I_{ab} \int dt \Omega^a \Omega^b ,\;\;\;
S_{\mathrm{trans}}[\bm{Z}] = \frac12 M_{\mathrm{cl}} \int dt
  \dot{\bm{Z}}\cdot  \dot{\bm{Z}}. 
\label{eq:30}
\end{align}
Here, $I_{ab}$ is the inertial tensor for the soliton and
$M_{\mathrm{cl}}$ is the mass of the classical soliton, which is found
  to be the sum of the $N_c$ valence-quark energies and the
  Dirac-continuum energy: $M_{\mathrm{cl}}=N_c E_{\mathrm{val}}+
  E_{\mathrm{sea}}$. We refer to Ref.~\cite{Kim:1995mr} for
  details. 

The integral over the translational zero modes yields naturally the
Fourier transform, which indicates that the baryon state has the
proper translational symmetry. Having performed the rotational
zero-mode quantization, we can restore the rotational symmetry so that
the baryon state has correct spin and flavor quantum numbers.  
After the zero-mode quantization, we obtain the collective Hamiltonian
as follows:
\begin{align}
H_{\mathrm{coll}} = H_{\mathrm{sym}} + H_{\mathrm{sb}},
\end{align}
where $H_{\mathrm{coll}}$ are decomposed into the flavor SU(3)
symmetric and symmetry-breaking terms 
\begin{align}
  \label{eq:Hamiltonian}
H_{\mathrm{sym}} = M_{\mathrm{cl}} + \frac1{2I_1} \sum_{i=1}^3
  \hat{J}_i^2 + \frac1{2I_2} \sum_{p=4}^7 \hat{J}_p^2,\quad
H_{\mathrm{sb}} = \alpha D_{88}^{(8)} + \beta \hat{Y} +
  \frac{\gamma}{\sqrt{3}} \sum_{i=1}^3 D_{8i}^{(8)} \hat{J}_i.
\end{align}
Here, $I_1$ and $I_2$ stand for the moments of inertia for the
soliton, which are the diagonal components of $I_{ab}$ in
Eq.~\eqref{eq:30} when $a=1,\,2,\,3$ and $a=4,\cdots,7$,
respectively. The explicit expressions for them can be found in
Appendix~\ref{app:A}. $D^{(8)}_{ab}$ represent SU(3) Wigner $D$ functions. 
The inertial parameters $\alpha$, $\beta$ and $\gamma$, which arise
from the linear $m_{\mathrm{s}}$ corrections, are expressed in terms
of the moments of inertia $I_1$ and $I_2$, and the anomalous moments
of inertia $K_1$ and $K_2$
\begin{align}
\alpha=\left (-\frac{{\Sigma}_{\pi N}}{3\overline{m}}
  +\frac{K_{2}}{I_{2}} \right )m_{\mathrm{s}},
  \quad \beta=-\frac{ K_{2}}{I_{2}}m_{\mathrm{s}}, 
  \quad \gamma=2\left ( \frac{K_{1}}{I_{1}}-\frac{K_{2}}{I_{2}}
  \right ) m_{\mathrm{s}},
\label{eq:alphaetc}  
\end{align}
where $\Sigma_{\pi N}$ stands for the pion-nucleon $\Sigma$ term and
its expression can be found in Appendix~\ref{app:A}. $K_1$ and $K_2$
arise from the rotation of the mass term $A^\dagger (\delta m) A$ in
Eq.~\eqref{eq:25} (see Ref.~\cite{Blotz:1992pw}). The corresponding
expressions can also be found in Appendix~\ref{app:A}.
Once the flavor SU(3) symmetry is broken, the collective wavefunctions
of the baryon decuplet start to get mixed with states in higher
representations. Thus, the states of the baryon octet and decuplet are
derived by the standard second-order perturbation theory: 
\begin{align}
|B_{{\bm{8}}_{1/2}}\rangle &= |{\bm{8}}_{1/2},B\rangle + 
  c^{B}_{\bm{\overline{10}}}|{{\overline{10}}}_{1/2},B\rangle + 
  c^{B}_{{27}}|{{\bm{27}}}_{1/2},B\rangle,
\label{eq:mixedWF1}
\\
|B_{{\bm{10}}_{3/2}}\rangle &= |{\bm{10}}_{3/2},B\rangle + 
  a^{B}_{{27}}|{{\bm{27}}}_{3/2},B\rangle + 
  a^{B}_{{35}}|{{\bm{35}}}_{3/2},B\rangle
\label{eq:mixedWF2}
\end{align}
with the mixing coefficients
\begin{eqnarray}
c_{\overline{10}}^{B}
\;=\;
c_{\overline{10}} \left[\begin{array}{c}
\sqrt{5}\\
0 \\
\sqrt{5} \\
0
\end{array}\right], 
& 
c_{27}^{B}
\;=\; 
c_{27}\left[\begin{array}{c}
\sqrt{6}\\
3 \\
2 \\
\sqrt{6}
\end{array}\right], 
\label{eq:pqmix1}
\end{eqnarray}

\begin{eqnarray}
a_{27}^{B}
\;=\;
a_{27}\left[\begin{array}{c}
\sqrt{15/2}\\
2 \\
\sqrt{3/2} \\
0
\end{array}\right], 
& 
a_{35}^{B}
\;=\; 
a_{35}\left[\begin{array}{c}
5/\sqrt{14}\\
2\sqrt{{5}/{7}} \\
3\sqrt{{5}/{14}} \\
2\sqrt{{5}/{7}}
\end{array}\right], 
\label{eq:pqmix2}
\end{eqnarray}
respectively, in the basis
$\left[N,\;\Lambda,\;\Sigma,\;\Xi \right]$ for the baryon octet and
$\left[\Delta,\;\Sigma^{*},\;\Xi^{*},\;\Omega\right]$ for the baryon
decuplet. The parameters $c_{\overline{10}}$, $c_{27}$, $a_{27}$ and $a_{35}$
are expressed in terms of $\alpha$ and $\gamma$. 
\begin{align}
c_{\overline{10}} \;=\;
{\displaystyle -\frac{I_{2}}{15} \left ( \alpha + \frac{1}{2}
  \gamma \right)},\;\;\; & 
c_{27} \;=\; {\displaystyle -\frac{I_{2}}{25} \left( \alpha 
  -\frac{1}{6}\gamma \right)}, 
\label{eq:pqmix3} \\
a_{27} \;=\;
{\displaystyle -\frac{I_{2}}{8} \left ( \alpha + \frac{5}{6}
  \gamma \right)}, \;\;\;& 
a_{35} \;=\; {\displaystyle -\frac{I_{2}}{24} \left( \alpha 
  -\frac{1}{2}\gamma \right)}. 
\label{eq:pqmix4}
\end{align} 
Each state in Eqs.~\eqref{eq:mixedWF1} and ~\eqref{eq:mixedWF2} is
given in terms of the SU(3) Wigner $D$ functions that
satisfy the quantization condition~\cite{Blotz:1992pw}.

The final expression for the axial-vector transition form factors is 
derived as 
$C_5^{A,{\mathrm{10}}\to 8}$ 
\begin{align}
C_{5}^{A, {10}\to {8}}(Q^{2}) =& \frac{ 
\langle D^{(8)}_{a3} \rangle}{3} 
  \{\mathcal{A}_{0}(Q^{2}) 
-\mathcal{A}_{2}(Q^{2})\} 
  +\frac{1}{3\sqrt{3} I_{1}} \left[ \langle D^{(8)}_{a8} \hat{J}_{3} 
  \rangle +\frac{2m_{\mathrm{s}}}{\sqrt{3}} K_{1} \langle D^{(8)}_{83} 
  D^{(8)}_{a8} \rangle \right]  
 \{\mathcal{B}_{0}(Q^{2})-\mathcal{B}_{2}(Q^{2})\} \cr  
& +\frac{d_{pq3}}{3 I_{2}} \left[ \langle D^{(8)}_{ap} \hat{J}_{q} 
  \rangle +\frac{2m_{\mathrm{s}}}{\sqrt{3}} K_{2} \langle D^{(8)}_{ap} 
  D^{(8)}_{8q} \rangle \right]
  \{\mathcal{C}_{0}(Q^{2}) -\mathcal{C}_{2}(Q^{2})\}   
  -\frac{i \langle D^{(8)}_{a3} \rangle}{6I_{1}} 
  \{\mathcal{D}_{0}(Q^{2})-\mathcal{D}_{2}(Q^{2})\} \cr 
& +\frac{2 m_{\mathrm{s}}}{9} ( \langle D^{(8)}_{a3} \rangle - \langle 
  D^{(8)}_{88}D^{(8)}_{a3} \rangle) \{\mathcal{H}_{0}(Q^{2})
  -\mathcal{H}_{2}(Q^{2})\}  -\frac{2 m_{\mathrm{s}}}{9} 
  \langle D^{(8)}_{83} D^{(8)}_{a8} \rangle 
  \{\mathcal{I}_{0}(Q^{2}) -    \mathcal{I}_{2}(Q^{2})\} \cr  
& -\frac{2 m_{\mathrm{s}}}{3\sqrt{3}} d_{pq3} \langle D^{(8)}_{ap} 
  D^{(8)}_{8q} \rangle \{\mathcal{J}_{0}(Q^{2})
  -\mathcal{J}_{2}(Q^{2})\},
\label{eq:C5tri} 
\end{align}
where $\langle \cdots \rangle$ represent the matrix
elements for the SU(3) Wigner $D$ functions between $B_8$ and $B_{10}$
collective states, which are expressed in terms of the SU(3)
Clebsch-Gordan coefficients. The results are explicitly given in
Appendix~\ref{app:B}. $\mathcal{A}_0(Q^2),\cdots, \mathcal{J}_2(Q^2)$
denote the Fourier transforms of the axial-vector transition
densities, which can be found in Appendix~\ref{app:A}.  

Since the matrix elements of the Wigner $D$ functions also contain the
linear $m_{\mathrm{s}}$ terms, the collective baryon states
get the linear $m_{\mathrm{s}}$ corrections from those in higher
representations. Thus, there are yet additional $m_{\mathrm{s}}$
corrections in addition to those shown in Eq.~\eqref{eq:C5tri}. Thus,
it is more convenient to decompose the contributions arising from
flavor SU(3) symmetry breaking into two terms 
\begin{align}
C_{5}^{A, {10}\to {8}} &= 
C_{5}^{A, {10}\to {8}(\mathrm{sym})}
+ C_{5}^{A, {10}\to {8}(\mathrm{op}) }
+ C_{5}^{A, {10}\to {8}(\mathrm{wf})}, 
\label{eq:C5_decompose}
\end{align}
where $C_{5}^{A, {10}\to {8}(\mathrm{sym})}$ denote the
contributions from the SU(3) symmetric part in Eq.~\eqref{eq:C5tri}
whereas $C_{5}^{A, {10}\to {8}(\mathrm{op})}$ and  
$C_{5}^{A, {10}\to {8}(\mathrm{wf})}$ come respectively from the
current-quark mass term in the effective chiral action~\eqref{eq:echa}
and from the collective wavefunctions. They are explicitly written as 
\begin{align}
C_{5}^{A, 10\to 8(\mathrm{sym})} &=
   \frac{\sqrt{5}}{90} \left( 
\begin{array}{c} 
2 \\ -T_{3} \\ -2T_{3} \\ \sqrt{3} 
\end{array} 
\right)
 \left[ 2(\mathcal{A}_{0} - \mathcal{A}_{2})
-  \frac{i(\mathcal{D}_{0}-\mathcal{D}_{2})}{I_{1}}\right]  
 -\frac{\sqrt{5}}{90}\left( 
\begin{array}{c} 
2 \\ -T_{3} \\ -2T_{3} \\ 1 
\end{array}
 \right)  
\frac{\mathcal{C}_{0} -\mathcal{C}_{2}}{I_{2}}  ,  
\label{eq:C53sym}
\\ 
C_{5}^{A, {10}\to {8}(\mathrm{op})} =& 
\frac{\sqrt{5}m_{\mathrm{s}}}{405} 
  \left\{ \left( 
\begin{array}{c} 
1 \\ -3T_{3} \\ -5T_{3} \\ \sqrt{3} 
\end{array} 
\right)
  \left[ \frac{K_{1}}{I_{1}} (\mathcal{B}_{0}  -\mathcal{B}_{2})
 -(\mathcal{I}_{0} -\mathcal{I}_{2}) \right]\right. \cr
& \hspace{0.8cm}\left. + \left( 
\begin{array}{c} 
7 \\ -3T_{3} \\ -8T_{3} \\ 4\sqrt{3} 
\end{array}
  \right) 
\left[\frac{K_{2}}{I_{2}} (\mathcal{C}_{0} - \mathcal{C}_{2}) 
-(\mathcal{J}_{0} -\mathcal{J}_{2}) \right] 
 - \left( 
\begin{array}{c} 4 \\ 0 \\ T_{3} \\ \sqrt{3} 
  \end{array} 
\right) (\mathcal{H}_0 -\mathcal{H}_2)\right\},
\label{eq:C53opcorr}\\
C_{5}^{A, {10}\to {8}(\mathrm{wf})} =& 
   \frac{\sqrt{15}}{1620}a_{27}\left\{ 
\left( 
\begin{array}{c} 
2\sqrt{2} \\ -3\sqrt{3}T_{3} \\ -7\sqrt{2}T_{3} \\ 0 
\end{array} 
\right) 
\left[ 2(\mathcal{A}_0   -\mathcal{A}_2)
-\frac{i(\mathcal{D}_0 -\mathcal{D}_{2})}{I_{1}} \right] 
+ 4\left( 
\begin{array}{c} \sqrt{10} \\ 0 \\ -\sqrt{2}T_{3} \\ 3
  \end{array} 
\right)
  \frac{(\mathcal{C}_{0}-\mathcal{C}_{2})}{I_{2}} \right\}\cr
&  -\frac{\sqrt{3}}{135} a_{35}
\left\{ \left( 
\begin{array}{c} 5\sqrt{2} \\ 0 \\ \sqrt{10}T_{3} \\ -3\sqrt{5}
  \end{array} 
\right) 
\left[ 2(\mathcal{A}_0  - \mathcal{A}_{2}) 
-\frac{i(\mathcal{D}_0  - \mathcal{D}_{2})}{I_{1}}
\right] - \left( 
\begin{array}{c} 
5\sqrt{2} \\ 0 \\ -\sqrt{10}T_{3} \\ 3\sqrt{5}
\end{array} 
\right) \frac{(\mathcal{C}_0   -\mathcal{C}_{2})}{I_{2}} \right\},
\label{eq:C53wfcorr}
\end{align}
where we have suppressed $Q^2$ dependence of $C_5^A$
and $T_{3}$ is the third component of the isospin operator.

Since we have assumed isospin symmetry, we can find the isospin
relations for the axial-vector transition form factors as
follows~\cite{Yang:2015era}:  
\begin{align}
(\Delta^{+} \rightarrow p) &=  (\Delta^{0} \rightarrow
                n)=-\frac{1}{\sqrt{3}}(\Delta^{++} \rightarrow p) 
=\frac{1}{\sqrt{3}}(\Delta^{-} \rightarrow n) = (\Delta^{0} \rightarrow p)
=-(\Delta^{+} \rightarrow n) \cr  
\sqrt{2}(\Sigma^{*+} \rightarrow \Sigma^{+})
&=-\sqrt{2}(\Sigma^{*-} \rightarrow \Sigma^{-})
=(\Sigma^{*0} \rightarrow
   \Sigma^{+})=(\Sigma^{*-}    \rightarrow \Sigma^{0}) 
=(\Sigma^{*+} \rightarrow \Sigma^{0})=(\Sigma^{*0} \rightarrow
                                     \Sigma^{-}) \cr 
(\Sigma^{*0} \rightarrow \Lambda) &=
    \frac{1}{\sqrt{2}}(\Sigma^{*-} \rightarrow \Lambda) 
=-\frac{1}{\sqrt{2}}(\Sigma^{*+} \rightarrow \Lambda) \cr
(\Xi^{*0} \rightarrow \Xi^{0})&=-(\Xi^{*-} \rightarrow \Xi^{-})
=\frac{1}{2}(\Xi^{*-} \rightarrow \Xi^{0}) =
\frac{1}{2}(\Xi^{*0} \rightarrow \Xi^{-}) \cr
(\Delta^{++} \rightarrow \Sigma^{+}) &=
    \frac{\sqrt{3}}{\sqrt{2}}(\Delta^{+} \rightarrow \Sigma^{0})
=\sqrt{3}(\Delta^{0} \rightarrow \Sigma^{-}) \cr
(\Sigma^{*0} \rightarrow p)&=\frac{1}{\sqrt{2}}
    (\Sigma^{*-} \rightarrow n) \cr
  (\Sigma^{*0} \rightarrow \Xi^{0})&=\sqrt{2}
    (\Sigma^{*0} \rightarrow \Xi^{-}) \cr 
  (\Xi^{*0} \rightarrow \Sigma^{+})&=
    \sqrt{2}(\Xi^{*-} \rightarrow \Sigma^{0}),
\label{eq:relationIsospin}
\end{align}
where $(B_{10}\to B_8)$ denote the axial vector form factors
$C_5^{A, B_{10}\to B_8}$. We also find several sum rules between
those form factors:
\begin{align}
(\Delta^{+} \rightarrow p) & =-\frac{1}{\sqrt{6}}
(\Xi^{*-} \rightarrow \Lambda)
-\frac{1}{\sqrt{2}}(\Xi^{*-} \rightarrow
   \Sigma^{0})+\frac{4}{\sqrt{3}}(\Sigma^{*0}
   \rightarrow \Lambda) 
+(\Xi^{*-} \rightarrow \Xi^{0}) \cr
(\Sigma^{*+} \rightarrow \Sigma^{+})& =
    -\frac{1}{\sqrt{3}}(\Sigma^{*0}
    \rightarrow \Lambda) 
+\frac{1}{\sqrt{6}}(\Xi^{*-} \rightarrow
   \Lambda)+\frac{1}{\sqrt{2}}(\Xi^{*-}
   \rightarrow \Sigma^{0})  \cr 
(\Sigma^{*0} \rightarrow \Lambda)&= \frac{\sqrt{3}}{\sqrt{2}}
(\Sigma^{*0} \rightarrow p)
+\frac{\sqrt{3}}{2\sqrt{2}}(\Xi^{*-} \rightarrow \Sigma^{0})
-\frac{1}{2\sqrt{2}}(\Xi^{*-} \rightarrow
  \Lambda)-\frac{\sqrt{3}}{2}(\Xi^{*-}
  \rightarrow \Xi^{0}) \cr 
(\Sigma^{*0} \rightarrow p)&=-\frac{1}{2}(\Xi^{*-} \rightarrow
                             \Sigma^{0}) 
+\frac{\sqrt{3}}{2}(\Xi^{*-} \rightarrow \Lambda)
+\frac{1}{\sqrt{6}}(\Omega^{-} \rightarrow \Xi^{0}) \cr
(\Xi^{*-} \rightarrow \Xi^{0})& = \frac{\sqrt{2}}{\sqrt{3}}
(\Xi^{*-} \rightarrow \Lambda)-\frac{\sqrt{2}}{\sqrt{3}}
(\Sigma^{*0} \rightarrow \Lambda) 
+\frac{1}{\sqrt{3}}(\Omega^{-} \rightarrow \Xi^{0})\cr
(\Delta^{+} \rightarrow \Sigma^{0})&=
-(\Xi^{*-} \rightarrow \Sigma^{0})+\frac{1}{\sqrt{3}}(\Xi^{*-}
                                     \rightarrow \Lambda). 
\end{align}

%%%%%%%%%%%%%%%%%%%%%%%
\section{Results and discussion}
\label{sec:4}
%%%%%%%%%%%%%%%%%%%%%%%
Before we compute the axial-vector transition form factors
of the baryon decuplet, we first discuss how the parameters are
fixed. In the $\chi$QSM, there are four different parameters: the
dynamical quark mass $M$, the cutoff mass $\Lambda$ in the
regularization functions, the strange current quark mass
$m_{\mathrm{s}}$, and the average of the up and down current quarks
$\overline{m}=6.131\, \mathrm{MeV}$, as mentioned in Section
III. $\overline{m}$ is determined by reproducing the physical value of
the pion mass, $m_\pi=140$ MeV. The strange current quark mass is
usually fixed by the kaon mass, $m_K = 495$ MeV. Its value is obtained
to be 150 MeV. However, we use a slightly larger value
$m_{\mathrm{s}}=180$ MeV, which describes the mass spectra of the
baryon octet and decuplet~\cite{Blotz:1992pw, Christov:1995vm}. The
cutoff mass $\Lambda$ is determined by the pion decay constant
$f_\pi=93$ MeV.  
On the other hand, the dynamical quark mass $M$ is a free parameter in
the $\chi$QSM but is also fixed by reproducing the electric charge
radius of the proton~\cite{Kim:1995mr}, i.e., the corresponding value
of $M$ is $M=420$ MeV. We use exactly the same values of these
parameters in the present work.  
As shown in Eq.~\eqref{eq:extC5}, $C_5^A$ involves the octet mass
$M_8$. The baryon masses in the $\chi$QSM also include the rotational
$1/N_c$ and $m_{\mathrm{s}}$ corrections. If we turn off all the
corrections, the baryon masses become the classical nucleon mass
$M_{\mathrm{cl}}$ or the soliton mass, which is proportional to
$N_c$. To be theoretically more consistent, we will take
$M_{\mathrm{cl}}$ instead of a octet baryon
mass~\cite{Meissner:1986js, Ledwig:2008es}. In fact, the 
numerical results are improved by considering $M_{\mathrm{cl}}$ in
place of $M_8$ by around 10~\%. Similar effects can be seen in the
calculation of the magnetic dipole moments of the SU(3) baryons. 

\begin{figure}[ht]
  \includegraphics[scale=0.55]{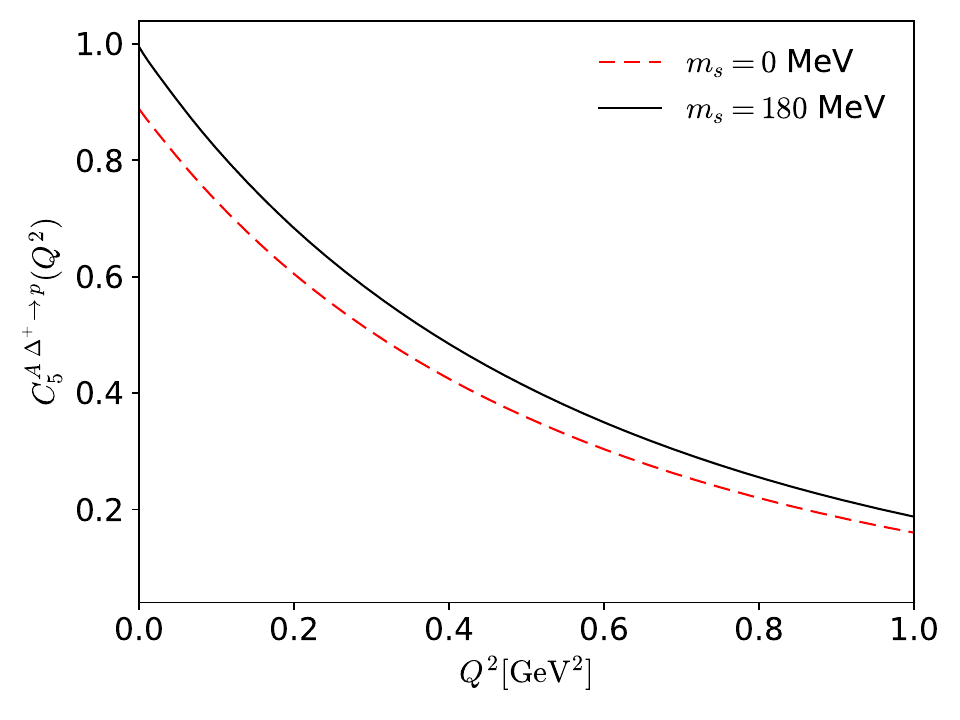}
  % \hspace{0.5cm}
\caption{Effects of the explicit flavor SU(3) symmetry breaking on  
the axial-vector transition form factors $C_{5}^{A, \Delta^{+}
  \rightarrow p}(Q^{2})$ for the $\Delta^+\to p$ transition. The solid
curve draws the total result whereas the dashed one depicts the result
without the $m_{\mathrm{s}}$ corrections.}
\label{fig:1}
\end{figure}
We first examine the effects of flavor SU(3) symmetry breaking on the
axial-vector transition form factor for the $\Delta^+\to p$
transition. In Fig.~\ref{fig:1}, we draw the results for the
$\Delta^+\to p$ axial-vector transition form factors. The solid curve
depicts the total result, whereas the dashed one draws that with the
effects of the explicit flavor SU(3) symmetry breaking turned off. The 
corrections from the linear $m_{\mathrm{s}}$ contribute to $C_{5}^{A,
  \Delta^{+} \rightarrow p}(Q^{2})$ by about 10~\%, as expected. As
discussed already in Ref.~\cite{Kim:2020lgp}, the effects of the
explicit flavor SU(3) symmetry breaking range in magnitude from
5 to 15~\%, depending on the decay modes. So, the linear
$m_{\mathrm{s}}$ corrections are also marginal in the case of the
$\Delta^+\to p$ axial-vector transition form factors. 

\begin{figure}[ht]
\includegraphics[scale=0.6]{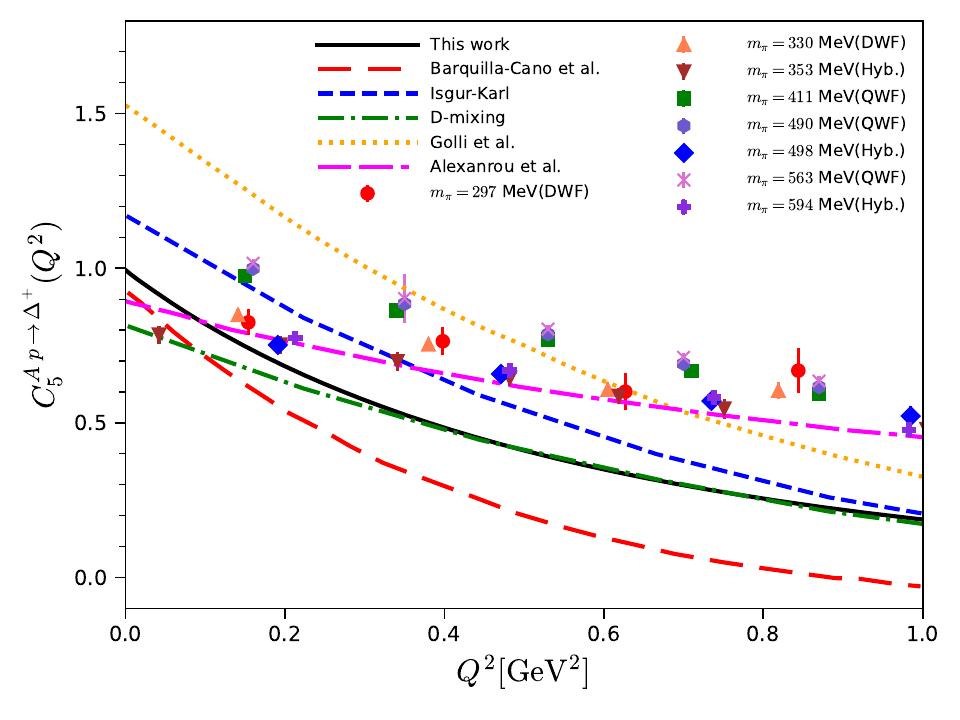}
\caption{Numerical results of the $C_{5}^{A(3)B_{10} \rightarrow
    B_{8}}(Q^{2})$ for the transition from the $\Delta^{+}$ isobar to
  the proton in comparison with those from other models. The solid
  curve draws the present result, whereas long-dashed, dashed,
  dot-dashed, dotted ones are taken from Refs.~\cite{Liu:1995bu,
    Golli:2002wy,   Barquilla-Cano:2007vds}. The present results are
  also compared with the data taken from lattice
QCD~\cite{Alexandrou:2007eyf,Alexandrou:2010uk}. The dot-dot-dashed
curve depicts a fit to a monopole form of the quenched lattice data
(see Fig.~17 in Ref.~\cite{Alexandrou:2007eyf}).}  
\label{fig:2}
\end{figure}
A few works computed theoretically the axial-vector transition form  
factors~\cite{Barquilla-Cano:2007vds, Liu:1995bu, Golli:2002wy}.  
So, we first compare the current result for $C_5^{A, \Delta\to N}
(Q^2)$ with those from other models as shown in Fig.~\ref{fig:2}. The
solid curve draws the present result, whereas the long-dashed one is
taken from Ref.~\cite{Barquilla-Cano:2007vds}, in which the chiral 
constituent quark model was used. In
Ref.~\cite{Barquilla-Cano:2007vds}, the effective Hamiltonian was 
constructed by introducing a confinement potential, a one-gluon
exchange potential, and a one-pion exchange 
potential. Because of the one-pion exchange potential, 
the model is called the chiral constituent quark model. Since the
nucleon and $\Delta$ states are constructed in terms of five harmonic
oscillator bases, the nonvalence-quark contributions are expressed by
states corresponding to $qqqq\bar{q}$ component. We want to mention
that the decomposition of the Fock space in quantum field theory can
only rigorously be performed in the light-cone
basis~\cite{Brodsky:1997de}.  Note that they use the empirical value 
of the axial transition mass $M_A\approx1.28\,\mathrm{GeV}$ as an
input, whereas it is predicted in the present work. The result of $C_5^A(0)$ in 
Ref.~\cite{Barquilla-Cano:2007vds} is completely determined by the
one-body axial-vector current, while the exchange-current
contributions are almost canceled by each other. Thus, the value of
$C_5^A(0)$ is obtained to be $C_5^A(0)=0.93$, which is very similar to
the present result: $C_5^A(0)=0.994$. On the other hand, the $Q^2$
dependence of $C_5^A$ from Ref.~\cite{Barquilla-Cano:2007vds} is 
quite different from the present one, as shown in Fig.~\ref{fig:2}.
That from Ref.~\cite{Barquilla-Cano:2007vds} falls off much faster
than the present result as $Q^2$ increases. 
The dashed and dot-dashed ones are obtained from
Ref.~\cite{Liu:1995bu}. Apart from the explicit 
forms of the potential, the model is similar to that used in
Ref.~\cite{Barquilla-Cano:2007vds}. In Ref.~\cite{Liu:1995bu}, three
different schemes were employed. The result in the short-dashed
curve, which was denoted by the Isgur-Karl (IK) model, was obtained by
using the parameters given in Refs.~\cite{Isgur:1977ef,
  Isgur:1978xj}. As shown in Fig.~\ref{fig:2}, the result of $C_5^A$
is larger and decreases faster than the present one as $Q^2$
increases. On the other hand, the result of $C_5^A(0)$ from the
$D$-state mixing model depicted in the dot-dashed curve is smaller
than the present one. However, its $Q^2$ dependence is milder than
that from the present one as well as that from the IK model. 
In Ref.~\cite{Golli:2002wy}, the linear sigma model and the cloudy bag
model were employed. The dotted curve in 
Fig.~\ref{fig:2} illustrates the result from the linear sigma model.
The value of $C_5^A(0)$ from Ref.~\cite{Golli:2002wy} is quite
overestimated in comparison with the fitted results from the T2K
experiment~\cite{T2K:2021xwb}.  

In Fig.~\ref{fig:2}, the dot-dot-dashed curve illustrates a fit to a
monopole form of the quenched lattice
data~\cite{Alexandrou:2007eyf,Alexandrou:2010uk}.  
It tends to fall off relatively slower than those of other models and
that of the present one. It is well known that the lattice 
calculations with the \emph{unphysical} pion mass produce in general 
hadronic form factors that fall off very slowly as $Q^2$
increases. Considering the picture that the pion fields govern the
structure of the nucleon and $\Delta$ in outer parts, one can
understand that the smaller pion mass renders the sizes of $N$ and
$\Delta$ smaller than physical ones. The result of the current work
for $C_5^{A, \Delta^+ \to p}$ is in good agreement with the lattice one
as will be shown explicitly in Table~\ref{tab:1}.

\begin{table}[htp]
\begin{threeparttable}
 \renewcommand{\arraystretch}{1.4}
{ \setlength{\tabcolsep}{21pt}
 \caption{Numerical results for the triplet axial-vector transition constant 
  $C^{A\ B_{10} \rightarrow B_{8}}_{5}(0)$ with $|\Delta S|=0$ in
  comparison with those from lattice QCD
  (LQCD)~\cite{Alexandrou:2010uk}, the relativistic quark  
  models (RQM)~\cite{Ravndal:1973xx, LeYaouanc:1976ea, Korner:1977rb},
  the isobar model~\cite{Fogli:1979cz}, the nonrelativistic quark
  model(NRQM)~\cite{Liu:1995bu},  the linear $\sigma$-model(LSM) and
  the cloudy bag model (CBM)~\cite{Golli:2002wy}, the chiral
  constituent quark model ($\chi$CQM)~\cite{Barquilla-Cano:2007vds},
  the relativistic baryon  chiral perturbation theory
  (RBCPT)~\cite{Geng:2008,Yao:2018pzc}, the Barbero-Lopez-Mariano
  approach ~\cite{Barbero:2008zza,Barbero:2013eqa}, Graczyk et al.'s
  work~\cite{Graczyk:2009qm}, Hernandez et al.'s
  work~\cite{Hernandez:2007qq}, the light-cone QCD 
  sum rule (LCSR)~\cite{Kucukarslan:2015urd}, and the nonlinear
  $\sigma$ model ~\cite{Alvarez-Ruso:2015eva}. We also 
compare the present results fitted to the T2K experimental
data(T2K)~\cite{T2K:2021xwb}.}   
 \label{tab:1}
 \begin{tabularx}{1.0\linewidth}{ c | c  c c c } %  l l l l l l
                                %  l l l l l   
  \hline 
  \hline 
$C^{A\ B_{10} \rightarrow B_{8}}_{5}(0)$ & $\Delta^{+} \rightarrow p$
 & $\Sigma^{*+} \rightarrow \Sigma^{+}$ & $\Sigma^{*0} \rightarrow \Lambda$ 
  & $\Xi^{*-} \rightarrow \Xi^{-}$ \\
  \hline 
$m_{\mathrm{s}} = 0\;\mathrm{MeV}$ 
& $0.888$ & $-0.443$ & $0.765$ & $0.412$\\
$m_{\mathrm{s}} = 180\;\mathrm{MeV}$ 
& $0.994$ & $-0.446$ & $0.840$ & $0.425$\\
\hline  
LQCD ~\cite{Alexandrou:2010uk}$(m_{\pi} = 297\;\mathrm{MeV})$ 
& $0.944 \pm 0.058^{*,\dagger}$ & -- & -- & --  \\
LQCD ~\cite{Alexandrou:2010uk}$(m_{\pi} = 330\;\mathrm{MeV})$ 
& $0.970 \pm 0.030^{*,\dagger}$ & -- & -- & -- \\
LQCD ~\cite{Alexandrou:2007eyf}$(m_{\pi} = 353\;\mathrm{MeV})$ 
& $0.750 \pm 0.019^{*,\dagger}$ & -- & -- & --  \\
LQCD ~\cite{Alexandrou:2007eyf}$(m_{\pi} = 411\;\mathrm{MeV})$ 
& $0.906 \pm 0.015^{*,\dagger}$ & -- & -- & --  \\
LQCD ~\cite{Alexandrou:2007eyf}$(m_{\pi} = 490\;\mathrm{MeV})$ 
& $0.930 \pm 0.014^{*,\dagger}$ & -- & -- & --  \\
LQCD ~\cite{Alexandrou:2007eyf}$(m_{\pi} = 498\;\mathrm{MeV})$ 
& $0.864 \pm 0.032^{*,\dagger}$ & -- & -- & --  \\
LQCD ~\cite{Alexandrou:2007eyf}$(m_{\pi} = 563\;\mathrm{MeV})$ 
& $0.952 \pm 0.016^{*,\dagger}$ & -- & -- & -- \\
LQCD ~\cite{Alexandrou:2007eyf}$(m_{\pi} = 594\;\mathrm{MeV})$ 
& $0.883 \pm 0.022^{*,\dagger}$ & -- & -- & -- \\
RQM1~\cite{Ravndal:1973xx} 
& $0.97$ & -- & --  & -- \\
RQM2~\cite{LeYaouanc:1976ea} 
& $0.83$ & -- & --  & -- \\
RQM3~\cite{Korner:1977rb} 
& $0.97$ & -- & --  & -- \\
Fogli et al.~\cite{Fogli:1979cz} 
& $1.18$ & -- & -- & -- \\
Liu et al.~\cite{Liu:1995bu} 
& $1.17$ & -- & -- & -- \\
LSM~\cite{Golli:2002wy} 
& $1.53$ & -- & -- & -- \\
CBM~\cite{Golli:2002wy} 
& $0.81$ & -- & -- & -- \\
$\chi $CQM~\cite{Barquilla-Cano:2007vds} 
& $0.93$ & -- & -- & -- \\
RBCPT1~\cite{Geng:2008} 
& $1.16$ & -- & -- & -- \\
Barbero et al.~\cite{Barbero:2008zza,Barbero:2013eqa} 
& $1.35$ & -- & -- & -- \\
Graczyk et al.~\cite{Graczyk:2009qm}
& $1.19\pm 0.08$ & -- & -- & -- \\
Hernandez et al.~\cite{Hernandez:2007qq}
& $0.867 \pm 0.075 $ & -- & -- & -- \\
LCSR~\cite{Kucukarslan:2015urd} 
& $1.14 \pm 0.20$ & -- & -- & -- \\
Alvarez-Ruso et al.~\cite{Alvarez-Ruso:2015eva} 
& $1.12 \pm 0.11$ & -- & -- & -- \\
RBCPT2~\cite{Yao:2018pzc} 
& $1.17 \pm 0.02$ & -- & -- & -- \\
T2K(Prefit)~\cite{T2K:2021xwb} 
& $0.96 \pm 0.15$ & -- & -- & -- \\
T2K(Postfit)~\cite{T2K:2021xwb} 
& $0.98 \pm 0.06$ & -- & -- & -- \\
 \hline 
 \hline
\end{tabularx}}
\begin{tablenotes}\footnotesize
\item[*] Since the expressions for the axial-vector transition
  constants in Ref.~\cite{Alexandrou:2010uk} are different from the
  present one by $-1$, we have considered this factor for comparison. 
\item[$\dagger$] In Ref.~\cite{Alexandrou:2007eyf}, these values
  are extrapolated ones obtained by using the dipole parametrization. 
\end{tablenotes}
\end{threeparttable}
\end{table}
In Table~\ref{tab:1}, we list the values of $C_5^A(0)$ for four 
different axial-vector transitions, with and without the effects of
explicit SU(3) symmetry breaking. One can quickly obtain the values of
$C_5^A$ for all other channels from the isospin relations given in
Eq.~\eqref{eq:relationIsospin}. Since there are many results for the
$\Delta^+\to p$ axial-vector transition derived from other
works, we compare the current results with them. As already
discussed in Fig.~\ref{fig:1}, the effects of the explicit SU(3)
symmetry breaking on the $\Delta\to N$ transition are about
10~\%. While the contribution of the linear $m_s$ corrections to 
$C_5^{A\,\Sigma^{*0}\to\Lambda}(0)$ is similar to that of the $\Delta
\to N$ transition, the effects of explicit SU(3) symmetry breaking are
almost negligible. The final result for $C_5^{A\,\Delta\to N}(0)$ is
obtained to be 0.994, which is in good agreement with the T2K
data~\cite{T2K:2021xwb}. Those from Refs.~\cite{Ravndal:1973xx,
  Korner:1977rb, Barquilla-Cano:2007vds, Golli:2002wy,
  Hernandez:2007qq} are also in good agreement with the T2K data. That
from Ref.~\cite{LeYaouanc:1976ea} is underestimated but those from 
Refs.~\cite{Fogli:1979cz, Liu:1995bu,  Golli:2002wy, Geng:2008,
  Barbero:2008zza,Barbero:2013eqa, Graczyk:2009qm,
  Kucukarslan:2015urd, Alvarez-Ruso:2015eva, Yao:2018pzc} yield larger
values than the fitted results from the T2K data.  

\begin{table}
\begin{threeparttable}
 \renewcommand{\arraystretch}{1.4}
 {\setlength{\tabcolsep}{18pt}
 \caption{Numerical results for the axial transition mass in comparison with the
   lattice data~\cite{Alexandrou:2010uk, Alexandrou:2007eyf}, 
    that extracted from the Argonne National Laboratory(ANL)
   data~\cite{Barish:1978pj, Radecky:1981fn}, CERN BEBC
   data~\cite{Aachen-Bonn-CERN-Munich-Oxford:1980iaz}, that from the
   Brookhaven National Laboratory(BNL) data~\cite{Kitagaki:1990vs,
     Graczyk:2009qm}, MiniBooNE data~\cite{MiniBooNE:2010bsu}, and T2K
   fitted results~\cite{T2K:2021xwb}. We also compare the present
   results with those from other works~\cite{Rein:1980wg,
     Lalakulich:2005cs, Alvarez-Ruso:2015eva}. We use the dipole-type
   form factor for parametrization A. Parametrization B corresponds to
   Alder's parametrization~\cite{Adler:1968tw}.}    
 \label{tab:2}
 \begin{tabularx}{1.0\linewidth}{ c | c  c c c c c} %  l l l l l l l l l l l  
  \hline 
  \hline 
$M_{A}$ [GeV]  & $\Delta^{+} \rightarrow p$ 
 & $\Sigma^{*+} \rightarrow \Sigma^{+}$ & $\Sigma^{*0} \rightarrow \Lambda$ 
& $\Xi^{*0} \rightarrow \Xi^{0}$  \\
  \hline 
Parametrization A
& $0.863$  & $1.03$ & $1.03$ 
& $1.35$ \\
Parametrization B
& $1.17$  & $1.32$ & $1.31$  
& $1.47$ \\
\hline
LQCD ~\cite{Alexandrou:2010uk}$(m_{\pi} = 297\;\mathrm{MeV})$(dipole)
& $1.699 \pm 0.170 $  & $-$ & $-$ & $-$  \\
LQCD ~\cite{Alexandrou:2010uk}$(m_{\pi} = 329\;\mathrm{MeV})$(dipole)
& $1.588 \pm 0.070 $  & $-$ & $-$ & $-$  \\
LQCD ~\cite{Alexandrou:2007eyf}$(m_{\pi} = 353\;\mathrm{MeV})$(dipole)
& $2.202 \pm 0.113 $  & $-$ & $-$ & $-$  \\
LQCD ~\cite{Alexandrou:2007eyf}$(m_{\pi} = 411\;\mathrm{MeV})$(dipole)
& $1.534 \pm 0.036 $  & $-$ & $-$ & $-$  \\
LQCD ~\cite{Alexandrou:2007eyf}$(m_{\pi} = 490\;\mathrm{MeV})$(dipole)
& $1.537 \pm 0.033 $  & $-$ & $-$ & $-$  \\
LQCD ~\cite{Alexandrou:2007eyf}$(m_{\pi} = 498\;\mathrm{MeV})$(dipole)
& $1.892 \pm 0.101 $  & $-$ & $-$ & $-$  \\
LQCD ~\cite{Alexandrou:2010uk}$(m_{\pi} = 563\;\mathrm{MeV})$(dipole)
& $1.544 \pm 0.032 $  & $-$ & $-$ & $-$  \\
LQCD ~\cite{Alexandrou:2007eyf}$(m_{\pi} = 594\;\mathrm{MeV})$(dipole)
& $1.924 \pm 0.085 $  & $-$ & $-$ & $-$  \\
Fogli et al.~\cite{Fogli:1979cz} 
& $0.75$ & -- & --  & --  \\
ANL~\cite{Barish:1978pj}
& $0.93 \pm 0.11 $  & $-$ & $-$ & $-$  \\
BEBC~\cite{Aachen-Bonn-CERN-Munich-Oxford:1980iaz}
& $0.85 \pm 0.10 $  & $-$ & $-$ & $-$  \\
Rein et al.~\cite{Rein:1980wg}
& $0.95 $ & $-$  & $-$ & $-$  \\
BNL~\cite{Kitagaki:1990vs}
& $1.28^{+0.08}_{-0.10} $  & $-$ & $-$ & $-$  \\
Lalakulich et al.~\cite{Lalakulich:2005cs}$^{c}$
& $1.05 $ & $-$ & $-$  & $-$  \\
Lalakulich et al.~\cite{Lalakulich:2005cs}$^{d}$
& $0.95 $ & $-$ & $-$ & $-$  \\
Hernandez et al.~\cite{Hernandez:2007qq}
& $0.985 \pm 0.082 $ & -- & -- & -- \\
Graczyk et al.~\cite{Graczyk:2009qm}
& $0.94 \pm 0.04 $ & $-$ & $-$ & $-$  \\
MiniBooNE~\cite{MiniBooNE:2010bsu}
& $1.35 \pm 0.17 $ & $-$ & $-$ & $-$  \\
Alvarez-Ruso et al.~\cite{Alvarez-Ruso:2015eva} 
& $0.954 \pm 0.063$ & -- & -- & -- \\
T2K(Prefit)~\cite{T2K:2021xwb}
& $1.20 \pm 0.03 $ & $-$ & $-$ & $-$  \\
T2K(Postfit)~\cite{T2K:2021xwb}
& $1.13 \pm 0.08 $ & $-$ & $-$ & $-$  \\
 \hline 
 \hline
\end{tabularx}}
\begin{tablenotes}\footnotesize
\item[c] They use the parametrization form as
  $C_5^A(Q^2)=\frac{C_{5}^{A}(0)}{(1+Q^{2}/M_{A}^{2})^{2}}
  \frac{1}{1+2Q^{2}/M_{A}^{2}}$.  
\item[d] They use the parametrization form as 
  $C_5^A(Q^2)=\frac{C_{5}^{A}(0)}{(1+Q^{2}/M_{A}^{2})^{2}}
  \left(\frac{1}{1+Q^{2}/3M_{A}^{2}}\right)^{2}$.
\end{tablenotes}
\end{threeparttable}
\end{table}
The axial-transition form factors can be parametrized in terms of
the axial transition mass $M_A$. Two different parametrizations are
used, i.e., dipole-type parametrization 
\begin{align}
C_5^A(Q^2) = \frac{C_{5}^{A}(0)}{(1+Q^2/M_A^2)^2}  
\end{align}
and the Adler's one:
\begin{align}
C_{5}^{A}(Q^{2}) =\frac{C_{5}^{A}(0)[1+a
  Q^{2}/(b+Q^{2})]}{(1+Q^{2}/M_{A}^{2})^{2}},   
\end{align}
where $a$ and $b$ are fixed respectively to be $a= -1.2$ and $b=2.0$.
We use both the parametrizations and call the first one 
parametrization A and the second one parametrization B. In
Table~\ref{tab:2}, we list the present results for the axial
transition mass in the case of the $\Delta S=0$ axial-vector transitions.
In general, the value of $M_A$ from parametrization A is smaller than
that from parametrization B. 
The lattice calculations use the
dipole-type parametrization, while many works employ
Adler's one. The present result (parametrization A) for the
$\Delta^+\to p$ transition is much smaller than those from lattice QCD.
One can easily understand this difference, since the results for
$C_5^{A,\Delta^+\to p}$ from the lattice data fall off much slower
than the present one. The result for $M_A(\Delta^+\to p)$ is in good
agreement with the fitted results from the T2K. However, there is a
caveat: if one computes the axial transition mean-squre radius for the
$\Delta^+\to p$ decay by using the dipole-type and Adler-type
parametrizations, we find $\langle
r^2\rangle_{\Delta^{+}p}=0.531\,\mathrm{fm}^2$ and $\langle
r^2\rangle_{\Delta^{+}p}=0.470\,\mathrm{fm}^2$, respectively. It
indicates that the dipole-type parametrization yields a closer value
of $\langle r^2\rangle_{\Delta^{+}p}$ to the model result (see
Table~\ref{tab:3}, where we give the following value: $\langle 
r^2\rangle_{\Delta^{+}p}=0.542\,\mathrm{fm}^2$).

\begin{table}
\begin{threeparttable}
 \renewcommand{\arraystretch}{1.4}
 {\setlength{\tabcolsep}{26pt}
 \caption{Numerical results for the axial transition radius in
   comparison with those from various approaches: baryon chiral
   perturbation theory(BCPT)~\cite{Zhu:2002kh,Unal:2021byi},the chiral   
constituent quark model~($\chi$CQM)~\cite{Barquilla-Cano:2007vds},
nonrelativistic quark potential model~\cite{Liu:1995bu} with two
different methods (Isgur-Karl and D-mixing), and lattice
QCD~\cite{Alexandrou:2007eyf}.}   
 \label{tab:3}
 \begin{tabularx}{1.0\linewidth}{ c | c c c c } %  l l l l l l l l l l l  
  \hline 
  \hline 
 $\langle r^{2}\rangle_{B_{10}B_8} $ [$\mathrm{fm}^{2}$]  & $\Delta^{+}
 \rightarrow p$ 
  & $\Sigma^{*+} \rightarrow \Sigma^{+}$ & $\Sigma^{*0} \rightarrow \Lambda$ 
  & $\Xi^{*0} \rightarrow \Xi^{0}$  \\
  \hline 
& $0.542$ & $0.452$ & $0.452$ & $0.345$ \\
\hline
BCPT1~\cite{Zhu:2002kh}
& $0.424-0.498$ & $-$ & $-$ & $-$ \\
BCPT2~\cite{Unal:2021byi}
& $0.345$ & $-$ & $-$ & $-$ \\
$\chi$CQM~\cite{Barquilla-Cano:2007vds}
& $0.59$ & $-$ & $-$ & $-$ \\
Isgur-Karl~\cite{Liu:1995bu}
& $0.32$ & $-$ & $-$ & $-$ \\
D-mixing~\cite{Liu:1995bu}
& $0.30$ & $-$ & $-$ & $-$ \\
Lattice QCD~\cite{Alexandrou:2007eyf}
& $0.18$ & $-$ & $-$ & $-$ \\
 \hline 
 \hline
 % \end{tabular}}
\end{tabularx}}
\end{threeparttable}
\end{table}
The mean-square radii for the $B_{10}\to B_8$ axial vector transitions
give information on the behaviors of the corresponding form factors in
the vicinity of $Q^2=0$, since they are defined by
\begin{align}
\langle r^2\rangle_{B_{10} B_8} = - \left. 6\frac{d C_5^A(Q^2) }{d
  Q^2} \right|_{Q^2=0}.  
\end{align}
Table~\ref{tab:3} lists the results for the $\langle
r^2\rangle_{B_{10}B_8}$. In the second column, we compare the current
result for $\langle r^2\rangle_{\Delta N}$ with those from other
works and found that the present result is in agreement with that
from the $\chi$CQM~\cite{Barquilla-Cano:2007vds} whereas it is smaller
than the other works. Note that as the strangeness $|S|$ increases,
the magnitudes of the axial transition radii are reduced. So, we have
the inequality relation
\begin{align}
  \label{eq:inequality}
  \langle r^2\rangle_{\Delta N} > \langle r^2\rangle_{\Sigma^* \Sigma} >
  \langle r^2\rangle_{\Xi^* \Xi} . 
\end{align}

\begin{figure}[htp]
  \includegraphics[scale=0.5]{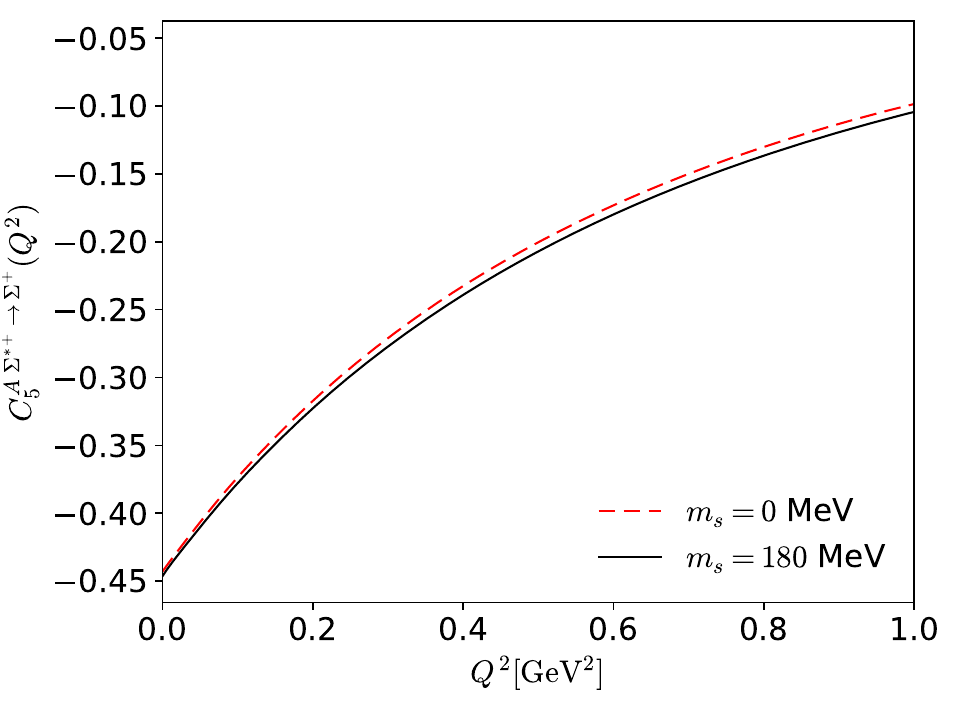} \;\;\;
  \includegraphics[scale=0.5]{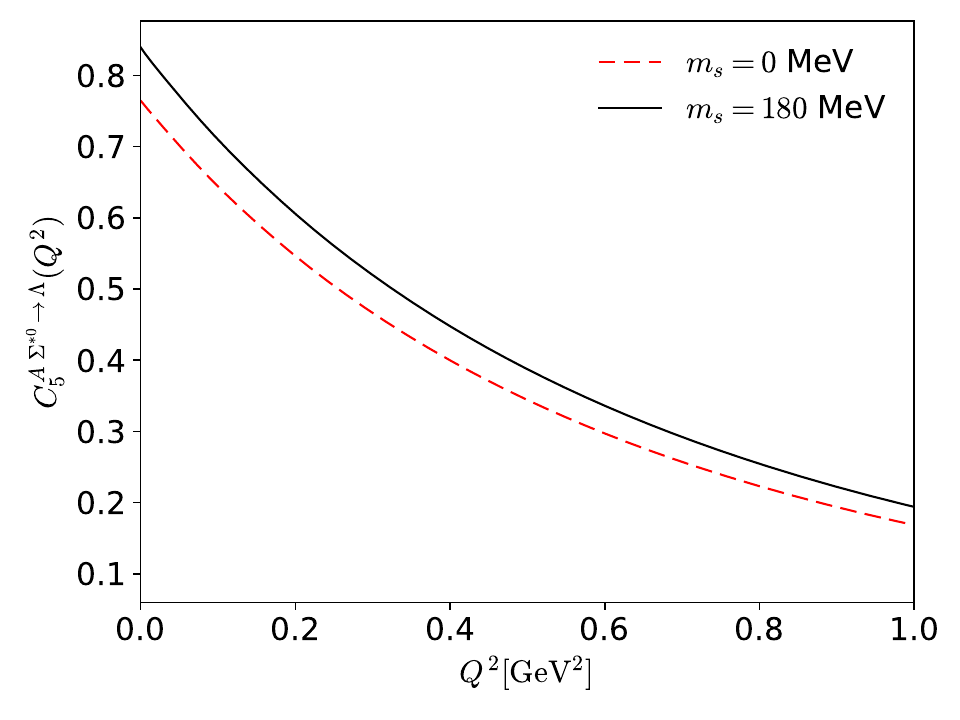}
\centering
\includegraphics[scale=0.5]{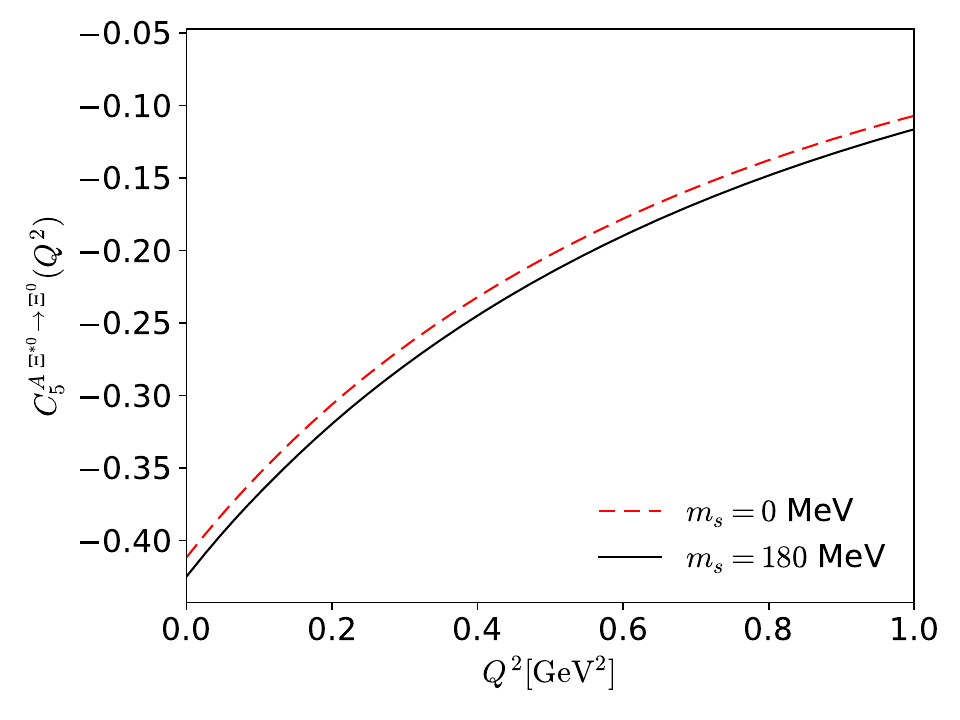}
\caption{Effects of the explicit flavor SU(3) symmetry breaking on
  $C_{5}^{A\,(3)\,\Sigma^{+*}\to \Sigma^+}(Q^{2})$ (upper left panel),
  $C_{5}^{A\,(3)\,\Sigma^{0*}\to \Lambda^0}(Q^{2})$ (upper right panel), and
  $C_{5}^{A\,(3)\,\Xi^{0*}\to \Xi^0}(Q^{2})$ (lower panel). Notations are
  the same as in Fig.~\ref{fig:1}.} 
  \label{fig:3}
\end{figure}

\begin{figure}[htp]
  \includegraphics[scale=0.55]{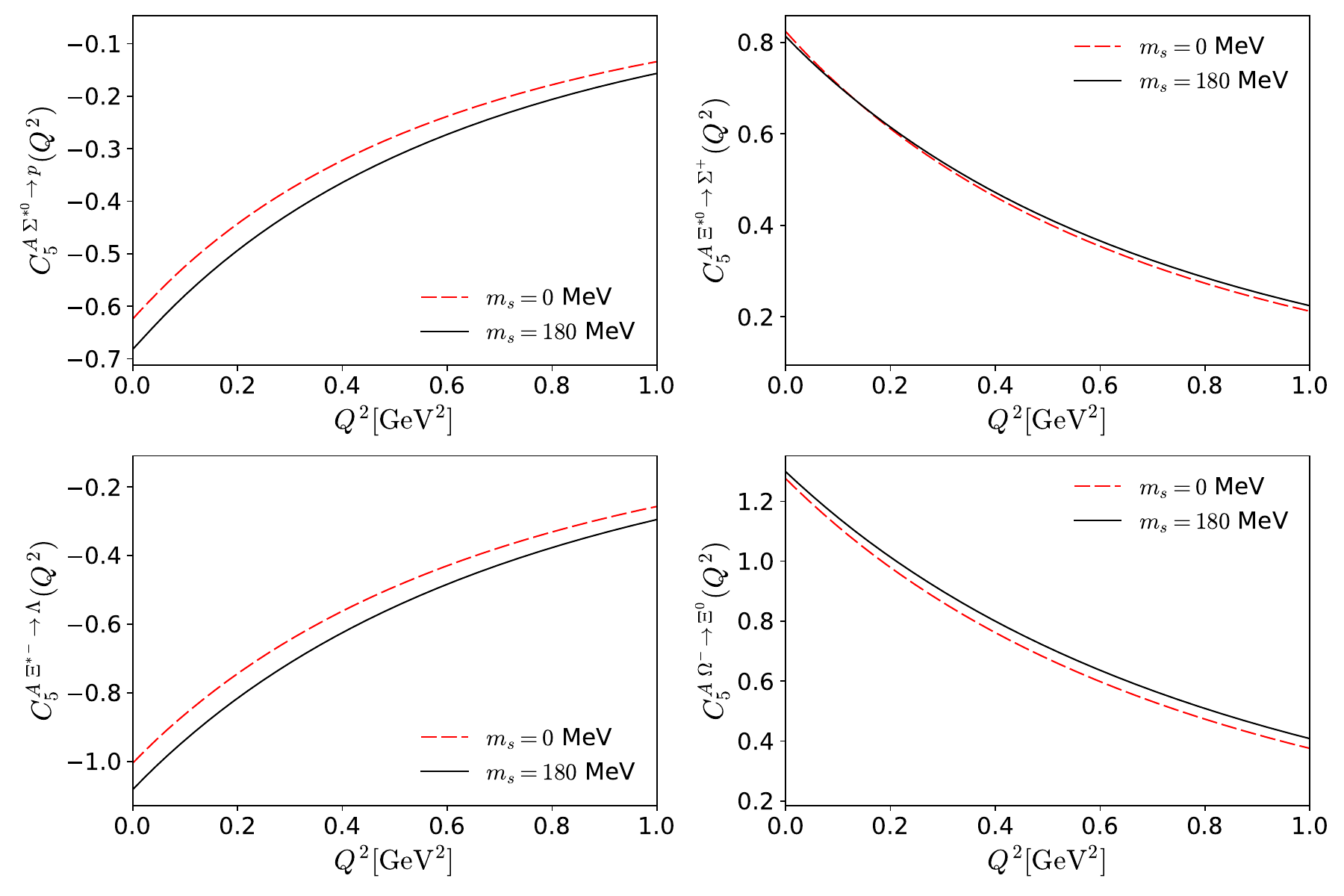}
    \includegraphics[scale=0.55]{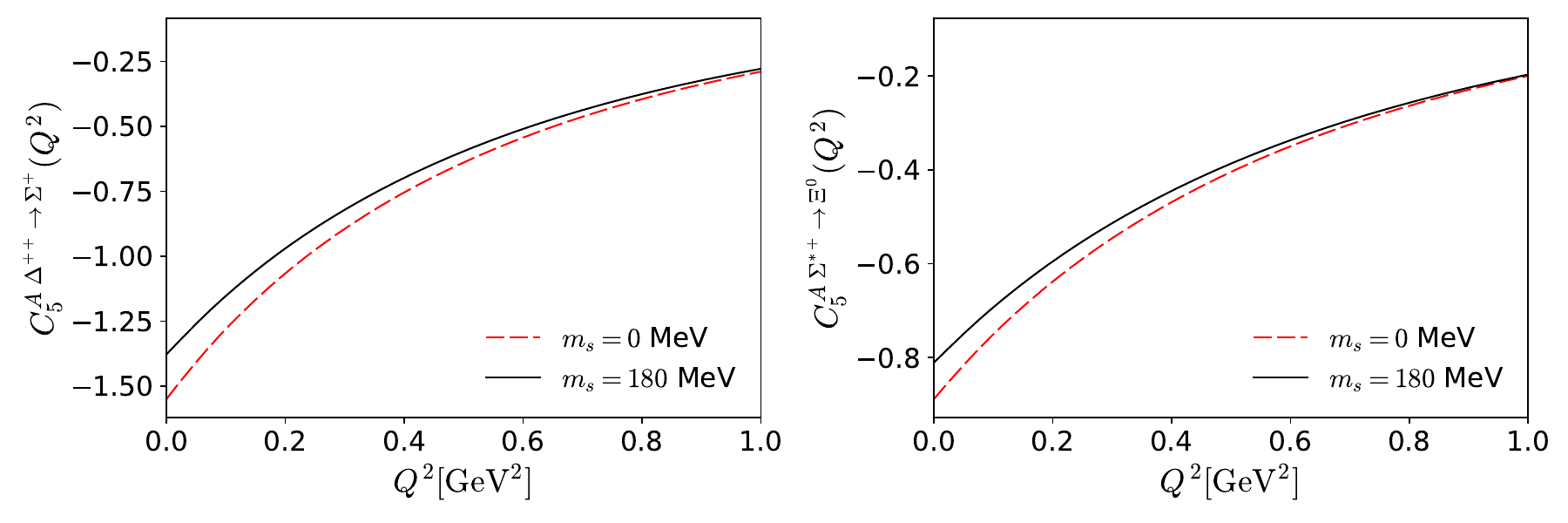}
\caption{Effects of the explicit flavor SU(3) symmetry breaking on
  $C_{5}^{A\,B_{10} \rightarrow B_{8}}(Q^{2})$ with $|\Delta S| = 1$. 
  Notations are the same as in Fig.~\ref{fig:1}.}
\label{fig:4}
\end{figure}

\begin{table}
\renewcommand{\arraystretch}{1.4}
{\setlength{\tabcolsep}{14pt}
\caption{Numerical results for $C^{A\ B_{10} \rightarrow
    B_{8}}_{5}(0)$ in comparison with those from the general framework
  of a chiral soliton model($\chi$SM) ~\cite{Yang:2015era}.
We use the dipole-type form factor for parametrization
A. Parametrization B corresponds to Alder's
parametrization~\cite{Adler:1968tw}.  
}
\label{tab:4}
 \begin{tabularx}{0.9\linewidth}{ c | c c c c}  
  \hline 
  \hline 
$C^{A\ B_{10} \rightarrow B_{8}}_{5}(0)$ & $\Sigma^{*0} \rightarrow p$  
 & $\Xi^{*0} \rightarrow \Sigma^{+}$ & $\Xi^{*-} \rightarrow \Lambda$ 
 & $\Omega^{-} \rightarrow \Xi^{0}$ \\
  \hline 
$m_{\mathrm{s}} = 0\;\mathrm{MeV}$ 
& $-0.624$  & $0.824$ & $-1.01$ & $1.28$ \\
$m_{\mathrm{s}} = 180\;\mathrm{MeV}$ 
& $-0.682$  & $0.813$ & $-1.08$ & $1.30$ \\
$\chi$SM\cite{Yang:2015era} 
& $-0.675 \pm 0.002$ & $0.954 \pm 0.003$ & $-1.169 \pm 0.004$ & $1.653
\pm 0.006$   \\ 
\hline 
$M_{A}$ [GeV] (A) 
& $1.25$ & $1.38$ & $1.37$ & $1.57$ \\
$M_{A}$ [GeV] (B)
& $1.32$ & $1.50$ & $1.49$ & $1.67$ \\
$\langle r^{2}\rangle $ [$\mathrm{fm}^{2}$](dipole)
& $0.375$ & $0.338$ & $0.342$ & $0.297$ \\ 
 \hline 
 \hline
 % \end{tabular}}
\end{tabularx}}
\end{table}

%p(4-i5)
\begin{table}
 \renewcommand{\arraystretch}{1.4}
{\setlength{\tabcolsep}{28pt}
\caption{Numerical results for $C^{A\ B_{10} \rightarrow
    B_{8}}_{5}(0)$ in comparison with those from the general framework
  of a chiral soliton model($\chi$SM) ~\cite{Yang:2015era}.
We use the dipole-type form factor for parametrization
A. Parametrization B corresponds to Alder's
parametrization~\cite{Adler:1968tw}.
} 
\label{tab:5}
 \begin{tabularx}{0.7\linewidth}{ c | c cc} 
 % \begin{tabular}{ c | c c c c c c} 
  \hline 
  \hline 
$C^{A\ B_{10} \rightarrow B_{8}}_{5}(0)$ & $\Delta^{++} \rightarrow
     \Sigma^{+}$ & $\Sigma^{*+} \rightarrow \Xi^{0}$ \\
  \hline 
$m_{\mathrm{s}} = 0\;\mathrm{MeV}$ 
& $-1.55$ & $-0.889$ \\
$m_{\mathrm{s}} = 180\;\mathrm{MeV}$ 
& $-1.38$ & $-0.811$ \\
$\chi$SM\cite{Yang:2015era} 
& $-1.547 \pm 0.006$ & $-0.928 \pm 0.004$ \\ 
\hline
$M_{A}$ [GeV] (A)
& $1.14$ & $1.27$ \\
$M_{A}$ [GeV] (B)
& $1.22$ & $1.37$ \\
$\langle r^{2}\rangle $ [$\mathrm{fm}^{2}$](dipole)
& $0.409$ & $0.368$ \\
 \hline 
 \hline
\end{tabularx}}
\end{table}
Figure~\ref{fig:3} draws the axial-vector transition form factors
$C_{5}^{A\,(3)\Sigma^{*+}\to \Sigma^+}(Q^{2})$,
$C_{5}^{A\,(3)\Sigma^{0*} \to \Lambda^0}(Q^{2})$, and
$C_{5}^{A\,(3)\Xi^{0*}\to \Xi^0}(Q^{2})$. The effects of flavor SU(3)
symmetry breaking on $C_5^{A\,(3)\Sigma^{*+} \to \Sigma^+}$ and
$C_5^{A\,(3)\Xi^{*0} \to \Xi^0}$ are neglibily small 
(below $4~\%$), whereas they contribute to
$C_{5}^{A\,(3)\Sigma^{0*}\to \Lambda^0}$ by about 10~\%. Thus, the
effects of flavor SU(3) symmetry breaking are overall marginal on the
axial-vector transition form factors. In Fig.~\ref{fig:4}, we
illustrate the axial-vector 
transition form factors $C_5^{A\, B_{10}\to B_8}$ with strangeness
changed. These transitions accompany the kaons in neutrino-nucleon
scattering to preserve strangeness. The results again show that the
effects of flavor SU(3) symmetry breaking contribute to $C_5^{A\,
  B_{10}\to B_8}$ at most by about 10~\%. In Tables~\ref{tab:4} and 
\ref{tab:5}, we list the numerical results for $C_5^{A \,B_{10}\to
  B_8}(0)$, the corresponding axial transition mass $M_A$ and axial
transition radii. We compare the results for $C_5^{A \,B_{10}\to
  B_8}(0)$ with those obtained from the chiral soliton
model~\cite{Yang:2015era}, where all the dynamical parameters given in
the present work were fixed by using the experimental data on hyperon
semileptonic decays~\cite{PDG}. The uncertainties of the results from 
Ref.~\cite{Yang:2015era} reflect the experimental errors. Except for
$C_5^{A \,\Sigma^{*0}\to p}(0)$, the current results are slightly 
underestimated but are qualitatively in agreement, compared with those
from Ref.~\cite{Yang:2015era}.  The results for the axial transition
radii indicate that as the strangeness $|S|$ increases, the values of
$\langle r^2\rangle_{B_{10}B_{8}}$ are lessened:
\begin{align}
  \label{eq:2}
  \langle r^2\rangle_{\Delta^{++} p} > \langle r^2\rangle_{\Sigma^{*0} p} >
  \langle r^2\rangle_{\Sigma^{*+} \Xi^0} > \langle
  r^2\rangle_{\Xi^{*0} \Sigma^+}> \langle r^2\rangle_{\Xi^{*-}
  \Lambda^0} > \langle r^2\rangle_{\Omega^- \Xi^0}.
\end{align}

\section{Summary and conclusion \label{sec:5}}
In the present work, we aimed at investigating the axial-vector
transition form factors for the transitions from the
baryon decuplet to the baryon octet within the framework of the SU(3)
self-consistent chiral quark-soliton model. We considered the
rotational $1/N_c$ corrections and the effects of flavor SU(3)
symmetry breaking, dealing with the strange current quark
perturbatively. We found that the linear $m_{\mathrm{s}}$ corrections
are marginal and even tiny to be neglected, depending on the
transition modes. We first compared the results for the axial-vector
$\Delta^+\to p$ transition form factors with those from lattice QCD
and other models and phenomenological analyses.
We obtained the axial-vector transition form factor for the $\Delta^+
\to p$ transition at $Q^2=0$ as $C_5^{A}(0)=0.994$. We derived the
axial transition mass with the dipole-type parametrization as
$M_A=0.863$ GeV whereas we got $M_A=1.17$ with Adler's
parametrization. $\Delta^+\to p$ transition form factor at $Q^2=0$ is
in good agreement with the lattice data and the fitted results from
the T2K data. Since the axial transition mass plays a critical role in
understanding the neutrino-proton interaction such as $\nu p\to \mu^-
p\pi^+$, we used the dipole-type and Adler's parametrizations for the
$\Delta^+\to p$ form factor.  We obtained $M_A=1.17$ GeV with Adler's
parametrization employed. This result is in good agreement with the
fitting of the T2K data. We then computed the radius squared for the
$\Delta^+\to p$ transition. The result is larger than those from other
works but is in agreement with that from
Ref.~\cite{Barquilla-Cano:2007vds}. We also obtained the axial-vector
form factors for other transition modes, including the
strangeness-changing transitions. We found that 
the values of the axial transition radii decrease as the strangeness
of the transition modes increases. So far, we are not able to conclude
whether this tendency is model-independent. 
One can extend the present theoretical framework to investigate the
axial-vector transition form factors of the singly heavy baryons. The
corresponding works are under way. 
%-------------------------------------------------

% -------------------------------------------------
\begin{acknowledgments}
%-------------------------------------------------
The present work was supported by Basic Science Research Program
through the National Research Foundation of Korea funded by the Korean
government (Ministry of Education, Science and Technology, MEST),
Grant-No. 2021R1A2C2093368 and 2018R1A5A1025563.
\end{acknowledgments}

\appendix
\section{Explicit expressions for the moments and anomalous of
  inertia, the $\pi N$ sigma term, and the form factors} \label{app:A}
In this Appendix, we present the explicit expressions for the moments
and anomalous moments of inertia, the $\pi N$ sigma term, and the
$Q^2$-dependent functions given in Eqs.~\eqref{eq:C5tri}.
The moments of inertia $I_{1}$, $I_{2}$ are expressed as 
\begin{align}
&I_{1} = N_{c}\delta^{ij}\left(\frac{1}{2}\sum_{\epsilon_{n} \neq
  \epsilon_{v}}\frac{1}{\epsilon_{n}-\epsilon_{v}} 
\langle v|\tau^{i}|n \rangle \langle n| \tau^{j}|v\rangle 
+\frac{1}{4}\sum_{n,m}^{n \neq m} \langle n|\tau^{i}|m \rangle \langle
  m| \tau^{j}|n\rangle\mathcal{R}_{3}(\epsilon_{n},\epsilon_{m})
  \right) \cr 
&I_{2} =
  N_{c}\left(\frac{1}{4}\sum_{\epsilon_{n^{0}}}
  \frac{1}{\epsilon_{n^{0}}-\epsilon_{v}}\langle 
  n^{0}|v \rangle\langle v|n^{0} \rangle 
+\frac{1}{4}\sum_{n,m^{0}}^{n \neq m^{0}}\langle m^{0}|n
  \rangle\langle
  n|m^{0}\rangle\mathcal{R}_{3}(\epsilon_{n},\epsilon_{m^{0}})
  \right),  
\end{align}
and the anomalous moments of inertia are written by 
\begin{align}
&K_{1} = N_{c}\delta^{ij}\left(\frac{1}{2}\sum_{\epsilon_{n} \neq
  \epsilon_{v}}\frac{1}{\epsilon_{n}-\epsilon_{v}} 
\langle v|\tau^{i}|n \rangle \langle n| \gamma^{0}\tau^{j}|v\rangle 
+\frac{1}{4}\sum_{n,m}^{n \neq m} \langle n|\tau^{i}|m \rangle \langle
  m|
  \gamma^{0}\tau^{j}|n\rangle\mathcal{R}_{5}(\epsilon_{n},\epsilon_{m})
  \right) \cr 
&K_{2} =
  N_{c}\left(\frac{1}{4}\sum_{\epsilon_{n^{0}}}
  \frac{1}{\epsilon_{n^{0}}-\epsilon_{v}}\langle 
  n^{0}|v \rangle\langle v|\gamma^{0}|n^{0} \rangle 
+\frac{1}{4}\sum_{n,m^{0}}^{n \neq m^{0}} \langle m^{0}|\gamma^{0}|n
  \rangle\langle
  n|m^{0}\rangle\mathcal{R}_{5}(\epsilon_{n},\epsilon_{m^{0}})
  \right). 
\end{align}
The $\pi N$ sigma term is expressed as
\begin{align}
&\Sigma_{\pi N} = -N_{c}\left(1-\frac{1}{\sqrt{3}}D_{88}^{(8)}\right)
\left[\langle v|\gamma^{0}|v \rangle +\sum_{n} \langle n|\gamma^{0}|n
  \rangle\mathcal{R}_{1}(\epsilon_{n}) -\mathrm{vacuum\,\,
  subtraction} 
  \right]. 
\end{align}  

$\mathcal{A}_0(Q^2)$, $\cdots$, $\mathcal{J}_0(Q^2)$
are defined by
\begin{align}
\mathcal{A}_{0}(Q^{2}) &=
 N_{c} \mathcal{M}
  \int d^{3} r j_{0}(Q |\bm{r}|) \left[ \phi^{\dagger}_{\mathrm{val}}
  (\bm{r}) \bm{\sigma} \cdot \bm{\tau} \phi_{\mathrm{val}}(r)
  + \sum_{n} \phi^{\dagger}_{n}(\bm{r}) \bm{\sigma} \cdot \bm{\tau} 
  \phi_{n}(\bm{r}) \mathcal{R}_{1}(E_n) \right] ,\cr
\mathcal{B}_{0}(Q^{2}) &=  N_{c}\mathcal{M} 
  \int d^{3} r j_{0}(Q |\bm{r}|) \left[ \sum_{n \ne
  \mathrm{val} } \frac{1}{E_{\mathrm{val}}-E_{n}} 
  \phi^{\dagger}_{\mathrm{val}}(\bm{r}) \bm{\sigma} 
  \phi_{n}(\bm{r}) \cdot \langle n | \bm{\tau} | \mathrm{val} \rangle 
  \right. \cr
& \left. \hspace{3.9cm} 
  -\frac{1}{2} \sum_{n,m} \phi^{\dagger}_{n}(\bm{r}) \bm{\sigma} 
  \phi_{m}(\bm{r}) \cdot \langle m | \bm{\tau} | n \rangle 
  \mathcal{R}_{5}(E_n,E_m) \right],\cr 
\mathcal{C}_{0}(Q^{2}) &=  N_{c} \mathcal{M}
  \int d^{3} r j_{0}(Q |\bm{r}|) \left[ \sum_{n_{0} \ne
  \mathrm{val} } \frac{1}{E_{\mathrm{val}}-E_{n_{0}}} 
  \phi^{\dagger}_{\mathrm{val}}(\bm{r}) \bm{\sigma} \cdot \bm{\tau} 
  \phi_{n_{0}}(\bm{r}) \langle n_{0} | \mathrm{val} \rangle \right. \cr
& \left. \hspace{3.9cm} 
  -\sum_{n,m_{0}} \phi^{\dagger}_{n}(\bm{r}) \bm{\sigma} \cdot \bm{\tau} 
  \phi_{m_{0}}(\bm{r}) \langle m_{0} | n \rangle 
  \mathcal{R}_{5}(E_n,E_{m_{0}}) \right], \cr
\mathcal{D}_{0}(Q^{2}) &= N_{c} \mathcal{M}
  \int d^{3} r j_{0}(Q |\bm{r}|) \left[ \sum_{n \ne
  \mathrm{val} } \frac{\mathrm{sgn}(E_{n})}{E_{\mathrm{val}}-E_{n}} 
  \phi^{\dagger}_{\mathrm{val}}(\bm{r}) (\bm{\sigma} \times \bm{\tau})
  \phi_{n}(\bm{r}) \cdot \langle n | \bm{\tau} | \mathrm{val} 
  \rangle \right. \cr
& \left. \hspace{3.9cm} 
  + \frac{1}{2} \sum_{n,m} \phi^{\dagger}_{n}(\bm{r}) \bm{\sigma} 
  \times \bm{\tau} \phi_{m}(\bm{r}) \cdot \langle m | \bm{\tau} 
  | n \rangle \mathcal{R}_{4}(E_n,E_m) \right],\cr
\mathcal{H}_{0}(Q^{2}) &= N_{c} \mathcal{M}
  \int d^{3} r j_{0}(Q |\bm{r}|) \left[ \sum_{n \ne
  \mathrm{val} } \frac{1}{E_{\mathrm{val}}-E_{n}} 
  \phi^{\dagger}_{\mathrm{val}}(\bm{r}) \bm{\sigma} \cdot \bm{\tau} 
  \langle \bm{r} | n \rangle \langle n | \gamma^{0}| \mathrm{val} 
  \rangle \right. \cr
& \left. \hspace{3.9cm} 
  + \frac{1}{2} \sum_{n,m} \phi^{\dagger}_{n}(\bm{r}) \bm{\sigma} \cdot 
  \bm{\tau} \phi_{m}(\bm{r}) \langle m | \gamma^{0} | n \rangle 
  \mathcal{R}_{2}(E_n,E_m) \right], \cr
\mathcal{I}_{0}(Q^{2}) &= N_{c} \mathcal{M}
  \int d^{3} r j_{0}(Q |\bm{r}|) \left[ \sum_{n \ne
  \mathrm{val} } \frac{1}{E_{\mathrm{val}}-E_{n}} 
  \phi^{\dagger}_{\mathrm{val}}(\bm{r}) \bm{\sigma} \phi_{n}(\bm{r}) 
  \cdot \langle n | \gamma^{0} \bm{\tau} | \mathrm{val} \rangle 
  \right. \cr
& \left. \hspace{3.9cm} +\frac{1}{2} \sum_{n,m} \phi^{\dagger}_{n}
  (\bm{r}) \bm{\sigma} \phi_{m}(\bm{r}) \cdot \langle m | \gamma^{0} 
  \bm{\tau} | n \rangle \mathcal{R}_{2}(E_n,E_m)\right],\cr
\mathcal{J}(Q^{2}) &= N_{c} \mathcal{M}
  \int d^{3} r j_{0}(Q |\bm{r}|) \left[ \sum_{n_{0} \ne
  \mathrm{val} } \frac{N_{c}}{E_{\mathrm{val}}-E_{n_{0}}} 
  \phi^{\dagger}_{\mathrm{val}}(\bm{r}) \bm{\sigma} \cdot \bm{\tau} 
  \phi_{n_{0}}(\bm{r}) \langle n_{0}| \gamma^{0} | \mathrm{val} \rangle 
  \right. \cr
& \left. \hspace{3.9cm} +N_{c} \sum_{n,m_{0}} \phi^{\dagger}_{n}(\bm{r}) 
  \bm{\sigma} \cdot \bm{\tau} \phi_{m_{0}}(\bm{r}) 
  \langle m_{0}| \gamma^{0} | n \rangle \mathcal{R}_{2}(E_n,E_{m_{0}}) 
  \right],
\label{AxComp10}
\end{align}
where $\mathcal{M}$ is defined by
\begin{align}
\mathcal{M} = \sqrt{\frac{3M_8}{E_8+M_8}}.  
\end{align}
The regularization functions are defined as 
\begin{align}
\mathcal{R}_{1}(E_{n}) &= \frac{-E_{n}}{2 \sqrt{\pi}} \int^{\infty}_{0}
  \phi(u) \frac{du}{\sqrt{u}} e^{-u E_{n}^{2}}, \cr
\mathcal{R}_{2}(E_{n},E_{m}) &= \frac{1}{2 \sqrt{\pi}} \int^{\infty}_{0}
  \phi(u) \frac{du}{\sqrt{u}} \frac{ E_{m} e^{-u E_{m}^{2}} 
  -E_{n}e^{-uE_{n}^{2}}}{E_{n} - E_{m}}, \cr
\mathcal{R}_{4}(E_{n},E_{m}) &= \frac{1}{2 \pi} \int^{\infty}_{0} du
  \, \phi(u) \int^{1}_{0} d\alpha e^{-\alpha uE^{2}_{m} 
  -(1-\alpha)uE^{2}_{n}} \frac{(1-\alpha)E_{n}-\alpha E_{m}}
  {\sqrt{\alpha(1-\alpha)}}, \cr
\mathcal{R}_{5}(E_{n},E_{m}) &=
  \frac{\mathrm{sgn}(E_{n})-\mathrm{sgn}(E_{m})}{2(E_{n}-E_{m})}.
\end{align}
Here, $|\mathrm{val}\rangle$ and $|n\rangle$ denote the quark states
in the valence and Dirac continuum with the corresponding
eigenenergies $E_{\mathrm{val}}$ and $E_n$ of the one-body Dirac 
Hamiltonian $h(U)$, respectively.

$\mathcal{A}_2 (Q^2)$, $\cdots$,$\mathcal{J}_2 (Q^2)$ are defined by 
\begin{align}
\mathcal{A}_{2}(Q^{2}) &= N_{c} \mathcal{M}
  \int d^{3} r j_{2}(Q |\bm{r}|) \left[ \phi^{\dagger}_{\mathrm{val}}
  (\bm{r}) \left\{ \sqrt{2\pi}Y_{2} \otimes \sigma_{1}\right\}_{1} \cdot 
  \bm{\tau} \phi_{\mathrm{val}}(\bm{r}) 
  +\sum_{n} \phi^{\dagger}_{n}(\bm{r}) \left\{ \sqrt{2\pi}Y_{2} 
  \otimes \sigma_{1}\right\}_{1} \cdot \bm{\tau}
  \phi_{n}(\bm{r}) \mathcal{R}_{1}(E_n) \right], \cr
\mathcal{B}_{2}(Q^{2}) &= N_{c}\mathcal{M}
  \int d^{3} r j_{2}(Q |\bm{r}|) \left[ \sum_{n \ne
  \mathrm{val} } \frac{1}{E_{\mathrm{val}}-E_{n}} 
  \phi^{\dagger}_{\mathrm{val}}(\bm{r}) \left\{ \sqrt{2\pi}Y_{2} \otimes 
  \sigma_{1}\right\}_{1} \phi_{n}(\bm{r}) \cdot 
  \langle n | \bm{\tau} | \mathrm{val} \rangle \right. \cr
& \left. \hspace{3.9cm} 
  -\frac{1}{2} \sum_{n,m} \phi^{\dagger}_{n}(\bm{r}) 
  \left\{ \sqrt{2\pi}Y_{2} \otimes \sigma_{1}\right\}_{1} 
  \phi_{m}(\bm{r}) \cdot \langle m | \bm{\tau} | n \rangle 
  \mathcal{R}_{5}(E_n,E_m) \right],\cr
\mathcal{C}_{2}(Q^{2}) &= N_{c}\mathcal{M}
  \int d^{3} r j_{2}(Q |\bm{r}|) \left[ \sum_{n_{0} \ne
  \mathrm{val} } \frac{1}{E_{\mathrm{val}}-E_{n_{0}}} 
  \phi^{\dagger}_{\mathrm{val}}(\bm{r}) \left\{ \sqrt{2\pi}Y_{2} \otimes 
  \sigma_{1}\right\}_{1} \cdot \bm{\tau} \phi_{n_{0}}(\bm{r}) 
  \langle n_{0} | \mathrm{val} \rangle \right. \cr
& \left. \hspace{3.9cm} 
  -\sum_{n,m_{0}} \phi^{\dagger}_{n}(\bm{r}) \left\{ \sqrt{2\pi}Y_{2} 
  \otimes \sigma_{1}\right\}_{1} \cdot \bm{\tau} 
  \phi_{m_{0}}(\bm{r}) \langle m_{0} | n \rangle 
  \mathcal{R}_{5}(E_n,E_{m_{0}}) \right], \cr
\mathcal{D}_{2}(Q^{2}) &= N_{c} \mathcal{M}
  \int d^{3} r j_{2}(Q |\bm{r}|) \left[ \sum_{n \ne
  \mathrm{val} } \frac{\mathrm{sgn}(E_{n})}{E_{\mathrm{val}}-E_{n}} 
  \phi^{\dagger}_{\mathrm{val}}(\bm{r}) \left\{ \sqrt{2\pi}Y_{2} \otimes 
  \sigma_{1}\right\}_{1} \times \bm{\tau} 
  \phi_{n}(\bm{r}) \cdot \langle n | \bm{\tau} | \mathrm{val} 
  \rangle \right. \cr
& \left. \hspace{3.9cm} 
  + \frac{1}{2} \sum_{n,m} \phi^{\dagger}_{n}(\bm{r}) 
  \left\{ \sqrt{2\pi}Y_{2} \otimes \sigma_{1}\right\}_{1} 
  \times \bm{\tau} \phi_{m}(\bm{r}) \cdot \langle m | \bm{\tau} 
  | n \rangle \mathcal{R}_{4}(E_n,E_m) \right],\cr
\mathcal{H}_{2}(Q^{2}) &= N_{c} \mathcal{M}
  \int d^{3} r j_{2}(Q |\bm{r}|) \left[ \sum_{n \ne
  \mathrm{val} } \frac{1}{E_{\mathrm{val}}-E_{n}} 
  \phi^{\dagger}_{\mathrm{val}}(\bm{r}) \left\{ \sqrt{2\pi}Y_{2} \otimes 
  \sigma_{1}\right\}_{1} \cdot \bm{\tau} \langle \bm{r} 
  | n \rangle \langle n | \gamma^{0}| \mathrm{val} \rangle \right. \cr
& \left. \hspace{3.9cm} 
  + \frac{1}{2} \sum_{n,m} \phi^{\dagger}_{n}(\bm{r}) 
  \left\{ \sqrt{2\pi}Y_{2} \otimes \sigma_{1}\right\}_{1} \cdot 
  \bm{\tau} \phi_{m}(\bm{r}) \langle m | \gamma^{0} | n \rangle 
  \mathcal{R}_{2}(E_n,E_m) \right],\cr
\mathcal{I}_{2}(Q^{2}) &= N_{c} \mathcal{M}
  \int d^{3} r j_{2}(Q |\bm{r}|) \left[ \sum_{n \ne
  \mathrm{val} } \frac{1}{E_{\mathrm{val}}-E_{n}} 
  \phi^{\dagger}_{\mathrm{val}}(\bm{r}) \left\{ \sqrt{2\pi}Y_{2} \otimes 
  \sigma_{1}\right\}_{1} \phi_{n}(\bm{r}) \cdot 
  \langle n | \gamma^{0} \bm{\tau} | \mathrm{val} \rangle \right. \cr
& \left. \hspace{3.9cm} +\frac{1}{2} \sum_{n,m} 
  \phi^{\dagger}_{n}(\bm{r}) \left\{ \sqrt{2\pi}Y_{2} \otimes 
  \sigma_{1}\right\}_{1} \phi_{m}(\bm{r}) \cdot \langle m | \gamma^{0} 
  \bm{\tau} | n \rangle \mathcal{R}_{2}(E_n,E_m)\right], \nonumber
  \end{align}
  \begin{align}
\mathcal{J}_{2}(Q^{2}) &= N_{c} \mathcal{M}
  \int d^{3} r j_{2}(Q |\bm{r}|) \left[ \sum_{n_{0} \ne
  \mathrm{val} } \frac{N_{c}}{E_{\mathrm{val}}-E_{n_{0}}} 
  \phi^{\dagger}_{\mathrm{val}}(\bm{r}) 
  \left\{ \sqrt{2\pi}Y_{2} \otimes \sigma_{1}\right\}_{1} \cdot \bm{\tau} 
  \phi_{n_{0}}(\bm{r}) \langle n_{0}| \gamma^{0} | \mathrm{val} \rangle 
  \right. \cr
& \left. \hspace{3.9cm}
  +N_{c} \sum_{n,m_{0}} \phi^{\dagger}_{n}(\bm{r}) 
  \left\{ \sqrt{2\pi}Y_{2} \otimes \sigma_{1}\right\}_{1} \cdot \bm{\tau} 
  \phi_{m_{0}}(\bm{r}) \langle m_{0}| \gamma^{0} | n \rangle 
  \mathcal{R}_{2}(E_n,E_{m_{0}}) \right].
\label{AxComp12}
\end{align}
\section{Matrix elements of the SU(3) Wigner ${D}$ function}\label{app:B} 
In Table~\ref{tab:6} to \ref{tab:11}, we list the results for the
matrix elements of the 
relevant collective operators, which are required for the calculation
of the axial-vector transition form factors.
\begin{table}[ht]
\setlength{\tabcolsep}{8pt}
\renewcommand{\arraystretch}{2.2}
  \caption{The matrix elements of the single and double Wigner $D$
    functions when $a=3$.} 
  \label{tab:6}
\begin{center}
\begin{tabularx}{0.6\linewidth}{ c | c c c c} 
% \begin{tabular}{ c | c c c c } 
 \hline 
  \hline 
$B_{10} \rightarrow B_{8}$ & $\Delta \rightarrow N$ & $\Sigma^{*}  \rightarrow \Sigma$ & 
$\Xi^{*} \rightarrow \Xi$ & $\Sigma^{*} \rightarrow \Lambda$ \\  
 \hline
$\langle B_{\bm{8}} |D^{(8)}_{33} | B_{\bm{10}}\rangle$  
& $\frac{2\sqrt{5}}{15} $ & $-\frac{\sqrt{5}}{15} T_{3}$
& $-\frac{2\sqrt{5}}{15} T_{3}$ & $\frac{\sqrt{15}}{15}$ \\  
$\langle B_{\bm{8}} |D^{(8)}_{38} \hat{J}_{3} | B_{\bm{10}}\rangle$  
& $0$ & $0$
& $0$ & $0$ \\  
$\langle B_{\bm{8}} |d_{bc3} D^{(8)}_{3b} \hat{J}_{c} 
| B_{\bm{10}}\rangle$  
& $-\frac{\sqrt{5}}{15}$ & $\frac{\sqrt{5}}{30} T_{3}$
& $\frac{\sqrt{5}}{15} T_{3}$ & $-\frac{\sqrt{15}}{30}$ \\    
$\langle B_{\bm{8}} |D^{(8)}_{83}D^{(8)}_{38} | B_{\bm{10}}\rangle$
& $\frac{\sqrt{5}}{90}$ & $-\frac{\sqrt{5}}{30} T_{3}$
& $-\frac{\sqrt{5}}{18} T_{3}$ & $\frac{\sqrt{15}}{90}$ \\  
$\langle B_{\bm{8}} |D^{(8)}_{88}D^{(8)}_{33} | B_{\bm{10}}\rangle$
& $\frac{2\sqrt{5}}{45}$ & $0$
& $\frac{\sqrt{5}}{90} T_{3}$ & $\frac{\sqrt{15}}{90}$ \\  
$\langle B_{\bm{8}} |d_{bc3} D^{(8)}_{8c}D^{(8)}_{3b}| 
  B_{\bm{10}} \rangle$  
& $\frac{7\sqrt{15}}{270}$ & $-\frac{\sqrt{15}}{90} T_{3}$ 
& $-\frac{4\sqrt{15}}{135} T_{3}$ & $\frac{2\sqrt{5}}{45}$  \\  
 \hline 
 \hline
% \end{tabular}
\end{tabularx}
\end{center}
\end{table}

\begin{table}[ht]
\setlength{\tabcolsep}{8pt}
\renewcommand{\arraystretch}{2.2}
  \caption{The transition matrix elements of the single Wigner $D$ function operators
  coming from the 27-plet component of the baryon wavefunctions when $a=3$.}
  \label{tab:7}
\begin{center}
\begin{tabularx}{0.6\linewidth}{ c | c c c c}  
% \begin{tabular}{ c | c c c c } 
 \hline 
  \hline 
$B_{10} \rightarrow B_{8}$ & $\Delta \rightarrow N$ & $\Sigma^{*} \rightarrow \Sigma$ & 
$\Xi^{*} \rightarrow \Xi$ & $\Sigma^{*} \rightarrow \Lambda$ \\
\hline
$\langle B_{\bm{27}} |D^{(8)}_{33} | B_{\bm{10}} \rangle$
& $\frac{\sqrt{30}}{135}$ & $-\frac{\sqrt{5}}{30}T_{3}$
&  $-\frac{7\sqrt{30}}{270} T_{3}$ &  $\frac{2\sqrt{15}}{135}$ \\
$\langle B_{\bm{27}} |D^{(8)}_{38}J_{3} | B_{\bm{10}}\rangle$  
& $0$ & $0$ & $0$ & $0$ \\
$\langle B_{\bm{27}} |d_{ab3}D^{(8)}_{3a}J_{b} | B_{\bm{10}}\rangle$ 
& $\frac{2\sqrt{30}}{135}$ &$-\frac{\sqrt{5}}{15} T_{3} $ 
& $-\frac{7\sqrt{30}}{135} T_{3} $ &$\frac{4\sqrt{15}}{135}$ \\
$\langle B_{\bm{8}} |D^{(8)}_{33} | B_{\bm{27}} \rangle$
& $\frac{2\sqrt{6}}{27}$ & $0$
&  $-\frac{2\sqrt{30}}{135}T_{3}$ &  $\frac{2\sqrt{15}}{45}$ \\
$\langle B_{\bm{8}} |D^{(8)}_{38}J_{3} | B_{\bm{27}}\rangle$  
& $0$ & $0$ & $0$ & $0$ \\
$\langle B_{\bm{8}} |d_{ab3}D^{(8)}_{3a}J_{b} | B_{\bm{27}}\rangle$ 
& $\frac{\sqrt{6}}{27}$ & $0$
& $-\frac{\sqrt{30}}{135}T_{3}$ & $\frac{\sqrt{15}}{45}$ \\
 \hline 
 \hline
% \end{tabular}
\end{tabularx}
\end{center}
\end{table}

\begin{table}[ht]
\setlength{\tabcolsep}{11pt}
\renewcommand{\arraystretch}{2.2}
  \caption{The matrix elements of the single and double 
  Wigner $D$ functions for $a=4+i5$.}
  \label{tab:8}
\begin{center}
\begin{tabularx}{0.6\linewidth}{ c | c c c}  
% \begin{tabular}{ c | c c c } 
 \hline 
  \hline 
$B_{10} \rightarrow B_{8}$ & $\Delta^{++} \rightarrow \Sigma^{+}$ &
 $\Delta^{+} \rightarrow \Lambda$ 
& $\Sigma^{*+} \rightarrow \Xi^{-}$ \\  
 \hline
$\langle B_{\bm{8}} |D^{(8)}_{p3} | B_{\bm{10}}\rangle$  
& $-\sqrt{\frac{2}{15}} $ & $0$
& $-\frac{\sqrt{2}}{3\sqrt{5}}$ \\  
$\langle B_{\bm{8}} |D^{(8)}_{p8} \hat{J}_{3} | B_{\bm{10}}\rangle$  
& $0$ & $0$ & $0$ \\
$\langle B_{\bm{8}} |d_{bc3} D^{(8)}_{pb} \hat{J}_{c} 
| B_{\bm{10}}\rangle$  
& $\frac{1}{\sqrt{30}}$ & $0$
& $\frac{1}{3\sqrt{10}}$ \\
$\langle B_{\bm{8}} |D^{(8)}_{83}D^{(8)}_{p8} | B_{\bm{10}}\rangle$
& $\frac{1}{6\sqrt{30}}$ & $0$
& $\frac{1}{9\sqrt{10}}$ \\
$\langle B_{\bm{8}} |D^{(8)}_{88}D^{(8)}_{p3} | B_{\bm{10}}\rangle$
& $\frac{1}{6\sqrt{30}}$ & $0$
& $\frac{1}{9\sqrt{10}}$ \\
$\langle B_{\bm{8}} |d_{bc3} D^{(8)}_{8c}D^{(8)}_{pb}| 
  B_{\bm{10}} \rangle$  
& $\frac{\sqrt{2}}{9\sqrt{5}}$ & $0$
& $\frac{1}{9\sqrt{30}}$ \\   
 \hline 
 \hline
% \end{tabular}
\end{tabularx}
\end{center}
\end{table}

\begin{table}[ht]
\setlength{\tabcolsep}{11pt}
\renewcommand{\arraystretch}{2.2}
  \caption{The transition matrix elements of the single Wigner $D$ functions
  coming from the 27-plet component of the baryon wavefunctions for $a=4+i5$.}
  \label{tab:9}
\begin{center}
\begin{tabularx}{0.6\linewidth}{ c | c c c}  
% \begin{tabular}{ c | c c c} 
 \hline 
  \hline 
$B_{10} \rightarrow B_{27}$ & $\Delta^{++} \rightarrow \Sigma^{+}$ &
  $\Delta^{+} \rightarrow \Lambda$ 
& $\Sigma^{*+} \rightarrow \Xi^{-}$ \\  
\hline
$\langle B_{\bm{27}} |D^{(8)}_{p3} | B_{\bm{10}} \rangle$
& $-\frac{1}{6\sqrt{30}}$ & $0$ 
& $-\frac{1}{9\sqrt{15}}$ \\
$\langle B_{\bm{27}} |D^{(8)}_{p8}J_{3} | B_{\bm{10}}\rangle$  
& $0$ & $0$ & $0$  \\
$\langle B_{\bm{27}} |d_{ab3}D^{(8)}_{pa}J_{b} | B_{\bm{10}}\rangle$ 
& $-\frac{1}{3\sqrt{30}}$ & $0$ 
& $-\frac{2}{9\sqrt{15}}$ \\
$\langle B_{\bm{8}} |D^{(8)}_{p3} | B_{\bm{27}} \rangle$
& $\frac{2}{9}$ & $0$
& $\frac{2\sqrt{2}}{9\sqrt{5}}$ \\
$\langle B_{\bm{8}} |D^{(8)}_{p8}J_{3} | B_{\bm{27}}\rangle$  
& $0$ & $0$ & $0$ \\
$\langle B_{\bm{8}} |d_{ab3}D^{(8)}_{pa}J_{b} | B_{\bm{27}}\rangle$ 
& $\frac{1}{9}$ & $0$
& $\frac{\sqrt{2}}{9\sqrt{5}}$ \\
 \hline 
 \hline
% \end{tabular}
\end{tabularx}
\end{center}
\end{table}

\begin{table}[ht]
\setlength{\tabcolsep}{5pt}
\renewcommand{\arraystretch}{2.2}
  \caption{The matrix elements of the single and double 
  Wigner $D$ functions for $a=4-i5$.}
  \label{tab:10}
\begin{center}
\begin{tabularx}{0.6\linewidth}{ c | c c c c c c}  
% \begin{tabular}{ c | c c c c c c} 
 \hline 
  \hline 
$B_{10} \rightarrow B_{8}$ & $\Sigma^{*0} \rightarrow p$
& $\Xi^{*0} \rightarrow \Sigma^{+}$
& $\Xi^{*-} \rightarrow \Lambda$ & $\Omega^{-} \rightarrow \Xi^{0}$ \\  
 \hline
$\langle B_{\bm{8}} |D^{(8)}_{\Xi^{-}3} | B_{\bm{10}}\rangle$  
& $\frac{1}{3\sqrt{5}}$
& $-\frac{\sqrt{2}}{3\sqrt{5}}$ 
& $\frac{1}{\sqrt{15}}$ & $-\sqrt{\frac{2}{15}}$ \\
$\langle B_{\bm{8}} |D^{(8)}_{\Xi^{-}8} \hat{J}_{3} | B_{\bm{10}}\rangle$  
& $0$ & $0$ & $0$ & $0$ \\
$\langle B_{\bm{8}} |d_{bc3} D^{(8)}_{\Xi^{-}b} \hat{J}_{c}| B_{\bm{10}}\rangle$
& $-\frac{1}{6\sqrt{5}}$
& $\frac{1}{3\sqrt{10}}$
& $-\frac{1}{2\sqrt{15}}$ & $\frac{1}{\sqrt{30}}$ \\
$\langle B_{\bm{8}} |D^{(8)}_{83}D^{(8)}_{\Xi^{-}8}| B_{\bm{10}}\rangle$
& $-\frac{1}{9\sqrt{5}}$
& $-\frac{1}{18\sqrt{10}}$
& $-\frac{1}{4\sqrt{15}}$ & $0$ \\
$\langle B_{\bm{8}} |D^{(8)}_{88}D^{(8)}_{\Xi^{-}3}| B_{\bm{10}}\rangle$
& $\frac{1}{18\sqrt{5}}$
& $\frac{1}{9\sqrt{10}}$
& $0$ & $\frac{1}{2\sqrt{30}}$ \\
$\langle B_{\bm{8}} |d_{bc3} D^{(8)}_{8c}D^{(8)}_{\Xi^{-}b}| B_{\bm{10}} \rangle$
& $-\frac{1}{9\sqrt{15}}$
& $\frac{\sqrt{5}}{18\sqrt{6}}$
& $-\frac{1}{12\sqrt{5}}$ & $\frac{1}{6\sqrt{10}}$ \\    
 \hline 
 \hline
% \end{tabular}
\end{tabularx}
\end{center}
\end{table}

\begin{table}[ht]
\setlength{\tabcolsep}{5pt}
\renewcommand{\arraystretch}{2.2}
  \caption{The transition matrix elements of the single Wigner $D$ functions
  coming from the 27-plet component of the baryon wavefunctions for $a=4-i5$.}
  \label{tab:11}
\begin{center}
\begin{tabularx}{0.6\linewidth}{ c | c c c c}  
% \begin{tabular}{ c | c c c c c c} 
 \hline 
  \hline 
$B_{10} \rightarrow B_{27}$ & $\Sigma^{*0} \rightarrow p$
& $\Xi^{*0} \rightarrow \Sigma^{+}$
& $\Xi^{*-} \rightarrow \Lambda$ & $\Omega^{-} \rightarrow \Xi^{0}$ \\ 
\hline
$\langle B_{\bm{27}} |D^{(8)}_{\Xi^{-}3} | B_{\bm{10}} \rangle$
& $-\frac{2\sqrt{2}}{9\sqrt{15}}$
& $\frac{\sqrt{2}}{9\sqrt{5}}$ 
& $-\frac{1}{3\sqrt{15}}$ & $\frac{1}{6\sqrt{5}}$ \\
$\langle B_{\bm{27}} |D^{(8)}_{\Xi^{-}8}J_{3} | B_{\bm{10}}\rangle$  
& $0$ & $0$ & $0$ & $0$ \\
$\langle B_{\bm{27}} |d_{ab3}D^{(8)}_{\Xi^{-}a}J_{b} | B_{\bm{10}}\rangle$ 
& $-\frac{4\sqrt{2}}{9\sqrt{15}}$
& $\frac{2\sqrt{2}}{9\sqrt{5}}$ 
& $-\frac{2}{3\sqrt{15}}$ & $\frac{1}{3\sqrt{5}}$ \\
$\langle B_{\bm{8}} |D^{(8)}_{\Xi^{-}3} | B_{\bm{27}} \rangle$
& $\frac{2}{9\sqrt{5}}$
& $\frac{2}{9\sqrt{15}}$
& $\frac{\sqrt{2}}{3\sqrt{5}}$ & $0$ \\
$\langle B_{\bm{8}} |D^{(8)}_{\Xi^{-}8}J_{3} | B_{\bm{27}}\rangle$  
& $0$ & $0$ & $0$ & $0$ \\

$\langle B_{\bm{8}} |d_{ab3}D^{(8)}_{\Xi^{-}a}J_{b} | B_{\bm{27}}\rangle$ 
& $\frac{1}{9\sqrt{5}}$
& $\frac{1}{9\sqrt{15}}$ 
& $\frac{1}{3\sqrt{10}}$ & $0$ \\ 
 \hline 
 \hline
% \end{tabular}
\end{tabularx}
\end{center}
\end{table}
%=================================================

\end{document}